%% file: article.tex
\documentclass[twocolumn,secnumarabic,amssymb, nobibnotes,aps,pra]{revtex4-1}
\usepackage[english]{babel}

\usepackage[normalem]{ulem}
\usepackage[dvipsnames]{xcolor}
\usepackage{float}
\usepackage{amsfonts}
\usepackage{amsmath}
\usepackage{graphicx}
\usepackage{color}
\usepackage{hyperref}

\usepackage{pgfplots}
\pgfplotsset{compat=newest}
\usepackage{svg}
\usepackage{xcolor}

\begin{document}

\title{Depletion-limited Kerr solitons in singly-resonant optical parametric oscillators}
% https://www.overleaf.com/project/635b8119fb75890b8fdbf156
% Limit imposed by depletion in parametrically driven soliton
\author{Carlos Mas Arabí$^{1,*}$, Nicolas Englebert$^1$, Pedro Parra-Rivas$^{1,2}$, Simon-Pierre Gorza$^1$, and François Leo$^1$}
\affiliation{$^1$OPERA-photonics$,$ Université Libre de Bruxelles$,$ 50 Avenue F. D. Roosevelt$,$ CP 194/5 B-1050 Bruxelles$,$ Belgium}
\affiliation{$^2$Dipartimento  di  Ingegneria  dell\textquotesingle Informazione, Elettronica  e  Telecomunicazioni,
Sapienza  Universit\`a  di  Roma, via  Eudossiana  18, 00184  Rome, Italy}
\affiliation{$^*$ \href{mailto:carlos.mas.arabi@ulb.be}{carlos.mas.arabi@ulb.be}  }

\begin{abstract}
We analyze the impact of pump depletion in the generation of cavity solitons in a singly-resonant parametrical oscillator that includes a $\chi^{(3)}$ nonlinear section. We find an analytical expression that provides the soliton existence region using variational methods, study the efficiency of energy conversion, and compare it to a driven Kerr resonator modeled by the Lugiato-Lefever equation. At high walk-off, solitons in singly-resonant optical parametric oscillators are more efficient than those formed in a Kerr resonator driven through a linear coupler.
\end{abstract}

\maketitle
%Temporal cavity solitons (CSs) are pulses propagating without distortion in an optical driven resonator~\cite{akhmediev_dissipative_2005}. In the same fashion as conservative solitons, CSs can be shaped by the balance between dispersion and nonlinearity \cite{grelu_dissipative_2012}. However, CSs must counteract the loss through an external driving source. 

%There are many different ways to furnish the energy to a dissipative system. In mode-locked lasers, the forcing is provided though an active media that, in the presence of a mechanism of saturable absortion, can lead to the stable formation of dissipative solitons. In fiber based lasers in the anomalous dispersion regime, the soliton is mostly shaped by the interaction between dispersion and nonlinearity, in a process called soliton mode-locking. The solitons in this case are well-described by the cubic-quintic Ginzbug-Landau equation, and the phase of the pulses is not fixed.

Temporal optical solitons are localized structures propagating without changing their shape formed through the balance between nonlinearity and dispersion~\cite{j_r_taylor_optical}. They appear in lossless media, where a family of states exists for a given set of external parameters. A more general type of solitons exists in the presence of losses and is commonly known as dissipative solitons (DS)~\cite{akhmediev_dissipative_2005}. They are self-organized structures that, in addition to having a balance between dispersion and nonlinearity, need compensation of losses through an active medium or external forcing. That double balance constraint the family of temporal solitons that exist in the lossless limit to a smaller subset that can even become a unique stable soliton for a set of given parameters~\cite{boardman_soliton-driven_2001}.

\begin{figure}[htbp]
    \centering
    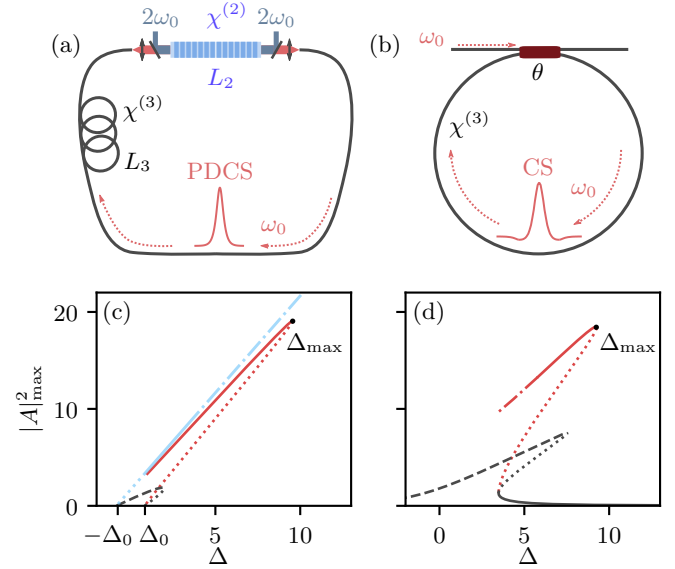
    \caption{Kerr resonators driven through parametric down-conversion (a) and a linear coupler (b). Stationary solutions for fixed values of driving of nonlinearly pumped  (c) and linearly driven (d) resonators. Black lines correspond to cw stationary solutions and red lines to solitons. The blue line in (c) corresponds to a nonlinear resonator with $\rho=0$. Solid lines represent stable solutions, dashed modulational unstable, dotted saddle unstable, and dashed-dotted Hopf unstable solutions.  Parameters (c) $\mu=1.3$, $\rho=0.32$, $d=6.25$. (d) $S=3.1$.}
    \label{fig:dispositif}
\end{figure}

An external driving can compensate for the system losses. The most common manner is to externally drive a temporal DS at frequency $\omega_0$  through a linear coupler. In a resonator with Kerr-type nonlinearity, DSs are known as Cavity Solitons (CSs)~\cite{wabnitz_suppression_1993}. In the temporal domain, CSs are pulses sitting atop a continuous wave (cw) background, while in the spectral domain, they consist in a $sech$ shape centered around the driving frequency ($\omega_0$) [see Fig. \ref{fig:dispositif} (b)]. They were first observed in a fiber-based cavity~\cite{leo_temporal_2010}, and have attracted a lot of attention since they underpin the formation of frequency combs in integrated platforms~\cite{herr_temporal_2014}, where they are sometimes called Dissipative Kerr Solitons (DKS). However, driving at $\omega_0$ through a linear coupler is inefficient, since only a small fraction of the pump energy is transferred to the CS~\cite{bao_nonlinear_2014}. 

An alternative way to drive a cavity is using a phase-sensitive nonlinear process. PDCSs can be shaped through quadratic nonlinearity~\cite{jankowski_temporal_2018,parra-rivas_dissipative_2022}, or Kerr effect~\cite{leonhardt_parametrically-driven_2022}. In our recent experiment, we  observed that Kerr CS can be parametrically driven using quadratic down conversion~\cite{englebert_parametrically_2021}, as theoretically predicted by Longhi~\cite{Longhi:95}. We coined this new type of solitons as parametrically driven cavity solitons (PDCS). Moreover, PDCSs have also been observed and studied in integrated platforms~\cite{bruch_pockels_2021,nie_deterministic_2021}. Parametric driving has been recently predicted to be potentially more efficient than driving through a linear coupler at $\omega_0$~\cite{li_break_2022}.
%In contrast to CS, PDCSs are backgrounless and have two different phases. This additional degree of freedom can be used for random number generation \cite{marandi_all-optical_2012} or Ising machines \cite{inagaki_large-scale_2016}.

Pump depletion sets a limit on the maximum attainable energy of PDCS. The depletion becomes relevant when the maximum power given to the soliton is not negligible compared to the pump power. Moreover, it plays a fundamental role in determining the frequency conversion efficiency from the driving to the PDCS. The impact of driving depletion depends on the system parameters and can be safely neglected in some regimes~\cite{englebert_parametrically_2021}. However, as we will show in this paper, it is generally not the case, and it limits the existence of PDCSs in OPOs. In this Letter, we present an analysis of PDCSs using  Lagrangian formalism. We develop an expression for their region of existence and confirm its validity using numerical path continuation. Using this analytical expression, we derive the efficiency of the soliton frequency conversion, which we compare with the one of an externally driven CS.

%These solitons have nothing to do with similaritons. Physically, the solitons that we are studying are Kerr solitons propagating in a cavity, whose losses are compensated by means of parametric amplification. 

% Fa falta que l'equacio (2) siga lequacio (1)

The cavity that we consider to generate PDCS is composed of two sections [see Fig. \ref{fig:dispositif}~(a)]: a quadratic material of length $L_2$ acting as a nonlinear coupler and a Kerr section of length $L_3$. The cavity is pumped through the quadratic material at $2\omega_0$. To model such a device, we use the normalized mean-field equation, including the Kerr effect:
\begin{equation}
    \partial_t A=-(1+i\Delta)A+i\partial_\tau^2A+i|A|^2A+B(t,\tau)A^*. 
    \label{eq:mean_field}
\end{equation}
$A=\sqrt{2\gamma L_3/\Lambda}E$ is the slowly varying envelope at $\omega_0$, where $\gamma$ is the third-order nonlinear coefficient, $\Lambda$ is the intensity loss coefficient, $L_3$ is the fiber length and $E$ is the slowly varying envelope with $\sqrt{W}$ units. $\tau=\left[\Lambda/(|\langle\beta_2(\omega_0)\rangle|L_3)\right]^{1/2} \bar{\tau}$ is the fast time, where $\langle\beta_2(\omega_0)\rangle$ is the averaged second-order dispersion at $\omega_0$ and $\bar{\tau}$ is the dimensional fast time; $t=\Lambda \bar{t}/(2T_c)$ is the slow time, where $\bar{t}$ has units of seconds and $T_c$ is the round-trip time.  

The terms on the right-hand side of Eq. (\ref{eq:mean_field}) describe, respectively losses,  detuning of the driving field from the closest resonance, the averaged dispersion of the cavity, the Kerr effect of the fiber, and the parametric gain. $\Delta=2\delta/\Lambda$ is the normalized detuning, where $\delta$ is the detuning expressed in radians. $B=\mu-\rho(A^2\otimes I)$ is the pump field at carrier frequency $2\omega_0$: The first term $\mu=2\kappa B_{in}L_2/\Lambda$ describes the cw component, where $\kappa$ is the second-order nonlinear coefficient, and $B_{in}$ is the input field. The second term in $B$ describes pump depletion through the convolution of $A$ with the kernel $I(\Omega)=(1-iD-e^{iD})/D^2$, and $D=-d\Omega-\eta_2 \Omega^2$ \cite{leo_walk-off-induced_2016,mosca_modulation_2018}. The kernel describes the depletion's frequency ($\Omega$) dependence and is a function of the crystal dispersion.  $\eta_2=\frac{\beta_2(2\omega_0)L_2\Lambda}{2\langle\beta_2(\omega_0)\rangle L_3}$ is the normalized group velocity dispersion at $2\omega_0$ in the quadratic material, where $\beta_2(2\omega_0)$ is the group velocity dispersion at $2\omega_0$, and $\langle\beta_2(\omega_0)\rangle$ is the averaged dispersion at $\omega_0$. The parameter $d=\Delta \beta_1 L_2\sqrt{\frac{\Lambda}{|\langle\beta_2(\omega_0)\rangle |L_3}}$ accounts  for the group velocity difference (walk-off) between pump and signal. We consider the half-frequency generation to be phase-matched. The coefficient $\rho=(\kappa L_2)^2/(\gamma L_3)$ is the relative contribution of $\chi^{(2)}$ with respect to the Kerr nonlinearity. 
%A complete derivation of Eq. (\ref{eq:mean_field}) can be found in the supplementary material. 

To show the impact of the pump depletion in a realistic device, we consider a resonator made of 20 m of single-mode fiber driven through a 4~cm  periodically poled lithium niobate crystal with a pump power of 170~mW at 775~nm. We suppose  5~\% losses per round trip, which corresponds to the following normalized parameters: $\mu=1.3$, where we considered $\kappa=2$~W$^{-1/2}$m$^{-1}$; $d=6.25$, taking $\langle\beta_2(\omega_0)\rangle=-10$~ps$^2$/km. $\eta_2$ is negligible with respect to $d$ in the Kernel, and thus the contribution of $\Omega^2$ in $I(\Omega)$ can be safely disregarded due to the limited bandwidth of the CS. 

Figure~\ref{fig:dispositif}~(c) shows the steady-state solutions in the slow time (i. e., $\partial_tA=0)$ of Eq. (\ref{eq:mean_field}) and their stability. For $\Delta<-\Delta_0=-\sqrt{1-\mu^2}$, the OPO exhibits a non-degenerated behavior that manifests as a modulational instability~\cite{englebert_parametrically_2021}. In the limit of negligible pump depletion ($\rho=0$), Eq.~(\ref{eq:mean_field}) is a parametrically driven nonlinear Schrödinger equation. It has analytical soliton solutions which read $A_{\text{sol}}=\sqrt{2}\beta\text{sech}(\beta \tau)e^{i\phi}$, where $\cos(2\phi)=\mu^{-1}$ and $\beta^2=\Delta+\mu\sin (2\phi)$~\cite{Longhi:95}. Solutions form two infinite branches that arise from $\Delta_0^2=1-\mu^2$ and extend to infinity~\cite{englebert_parametrically_2021}. The one arising from $-\Delta_0$ corresponds to stable PDCS [see Fig. \ref{fig:dispositif}~(c)].  The soliton energy  increases with $\Delta$, therefore, neglecting pump depletion is not valid at high detuning. Pump depletion ($\rho\neq0)$ limits the extension in detuning of the branches [see $\Delta_{\text{max}}=9.55$ in Fig. \ref{fig:dispositif}~(c)]. When $\rho\neq 0$, PDCSs do not have an analytical expression, and numerical methods are needed. We obtain the exact solutions by means of numerical continuation based on a standard Newton-Rapson method~\cite{allgower_introduction_2003}. We calculate their stability by numerically obtaining the eigenvalues of the Jacobian matrix associated to Eq.~(\ref{eq:mean_field})~\cite{skryabin_instabilities_1999}. 
 
 The maximum detuning reached by solitons ($\Delta_{\text{max}}$) depends on $\rho$, $\mu$ and $d$. Solitons are unstable from $\Delta_0$ up to $\Delta_{\text{max}}$, where they become stable in a saddle-node bifurcation. At lower $\Delta$, the branch connects to a series of unstable solutions that we do not study in this Letter.  It is interesting to compare the bifurcation diagram of a PDCS and a CS [Fig. \ref{fig:dispositif}~(d)] obtained in a driven Kerr cavity [Fig. \ref{fig:dispositif}~(b)], since both are Kerr solitons with similar intensity profiles. To model a linearly driven Kerr resonator, we use a standard Lugiato-Lefever Equation (LLE)~\cite{Coen_13_universal}. This equation corresponds to Eq. (\ref{eq:mean_field}), but replacing the term associated to the parametric driving (last term on the right-hand side of the equation) by a coherent driving $S$. To better compare the two systems, we set $S$ to match the same $\Delta_{\text{max}}$ as in the PDCS case ($S=3.1$).  At high detuning, the CS background is negligible since we are far from the resonance. Note that, within this limit, CS and PDNLS have approximately the same peak power [see Fig. \ref{fig:dispositif}~(c) and (d)]. Such a similarity will allow comparing the efficiency of PDCSs and  CSs.

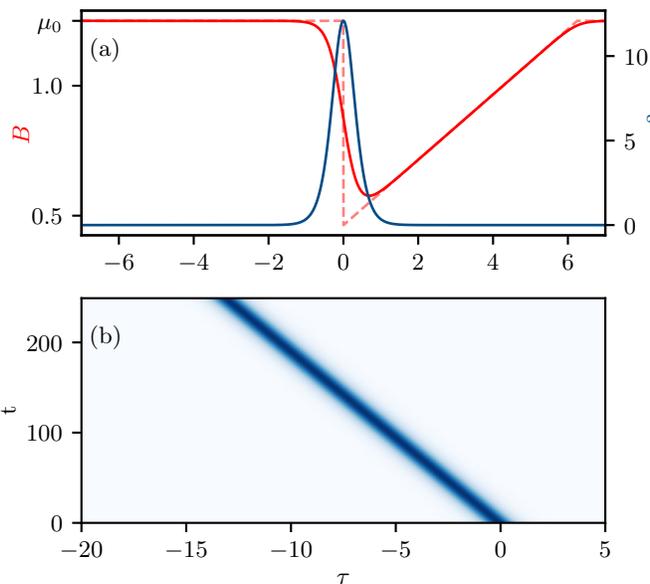
\begin{figure}
    \begin{center}
        \input{figura_3.pgf}
    \end{center}
    \caption{(a) Intensity of the pump field $B$(red). Dashed (solid) blue line shows the analytic approximation (numerically calculated depletion). Soliton with $\Delta=3$. (b) Evolution of a soliton in presence of pump depletion. ($\mu=1.1$, $d=6.25$, $\rho=0.2$)  }
    \label{fig:depletion}

\end{figure}

Figure \ref{fig:depletion}~(a) shows the pump field $B=\mu-\rho A^2\otimes I(\tau)$ with $d=6.25$. We consider that the soliton is short with respect to $I(\tau)$; hence, to calculate the convolution, the soliton can be approximated by a Dirac delta function. The depletion is approximated by the following analytical expression:
\begin{equation}
|A^2(\tau)\otimes I(\tau)|\approx \frac{2\beta}{|d|}\left(1-\frac{\tau}{d}\right)\left[\text{sign}\left(1-\frac{\tau}{d}\right)+\text{sign}\left(\frac{\tau}{d}\right)\right].
\label{eq:convolution}
\end{equation}
 The depletion amplitude at $\tau=0$ is inversely proportional to $|d|$. Then, the consequence  of a non-negligible walk-off is to spread the depletion in time and, as we will show later, to allow reaching shorter and more energetic solitons. Notice that the soliton takes energy from the pump $B$ in a region that extends between $\tau=0$ to $\tau=d$. Pump depletion also produces an asymmetry in the parametric amplification of the soliton and as a consequence,  leads to a temporal drift of the PDCS with respect to the reference frame traveling at the group velocity at $\omega_0$~[see Fig. \ref{fig:depletion}~(b)]. For $d>0$, the soliton has a higher group velocity than the reference frame due to a  more efficient amplification of the leading edge than the trailing one. In contrast, if $d<0$, the response is mirror symmetric, and the pulse travels slower than the reference frame. 
 
 %Note that here in this regime, in contrast to other structures formed in doubly  resonant degenerated OPOs such as simultons \cite{jankowski_temporal_2018} or walk-off induced solitons \cite{roy_temporal_2022}, the pump only amplifies the soliton, which does not give back energy through second harmonic generation to the pump.    

To find how $\Delta_{\text{max}}$ depends on $\rho$, $\mu$ and $d$  analytically, we use a variational  method. Eq. (\ref{eq:mean_field}) is a non-conservative equation. Therefore, we must modify the Euler-Lagrange equations to account for the system's dissipative part. Following a similar approach to ~\cite{yi_theory_2016,grelu_book}, and including amplification and loss through the Rayleigh functional ($R$), the modified Euler-Lagrange equations are

%% pose coses amb unitats

\begin{equation}
\frac{\partial L}{\partial x_k}-\frac{d}{dt}\frac{\partial L }{\partial x_k'}=\int (R^*\partial_{x_k'} A+R\partial_{x_k'} A^* )\text{d}\tau,
\label{eq:general}
\end{equation}
where $x_k$ are the different collective coordinates that describe the pulse,  and prime denotes $t$ derivative. The Lagrangian ($L$) is defined as $L=\int \mathcal{L} \text{d}\tau $, where $\mathcal{L}=\frac{i}{2}\left(A^*\partial_tA-A\partial_tA^*\right)-|\partial_\tau A|^2+\frac{|A|^4}{2}-\Delta |A|^2$ is the Lagrangian density of the nonlinear Schrödinger equation with anomalous dispersion~\cite{agrawal_nonlinear_2013}.  The perturbation $R$ that accounts for the dissipative terms reads
\begin{equation}
R=-i A+i\mu A^*-i\rho (A^2\otimes I)A^*.
\label{eq:R}
\end{equation}
 Considering that the soliton shape does not change with respect to the undepleted limit,  we choose as an \emph{ansatz} for the variational calculation  $A_{var}=\sqrt{2}\beta(t)\text{sech}[\beta(t)(\tau-\tau_0(t))]e^{i[\phi(t)-\Omega(t)(\tau-\tau_0(t))]}$, where the amplitude or temporal duration $\beta$, position $\tau_0$, phase $\phi$ and central frequency $\Omega$ are functions that depend on $t$. To reduce the number of free parameters in the $ansatz$, we have considered the soliton time duration and amplitude to be related in the same fashion as conservative Kerr solitons. Then the Lagrangian is
 \begin{equation}
L=-4\beta (\phi'+\Omega \tau_0')-4\beta \Omega^2+\frac{4}{3}\beta^3-4\Delta \beta.
\end{equation}
 Using the analytical approximation of the convolution [Eq. (\ref{eq:convolution})], the evolution equations of the collective variables are:
\begin{align}
\tau_0'&=\frac{-4\rho }{d}\left(\text{ln}(2)-\frac{\pi^2}{12 |d|\beta}\right), \nonumber \\
\beta'&=\beta\left[\mu\cos(2\phi)-1-\frac{2\beta\rho}{|d|}\left(1-\frac{\mathrm{ln}(2)}{\beta|d|}\right)\right], \nonumber \\
\phi'&=\beta^2-\Delta-\mu\sin(2\phi), \label{eq:dynamical_system}
\end{align}
where we focus on the solutions that satisfy $\Omega'=\Omega=0$. Solitons correspond to the fixed points of this dynamical system [Eqs.~(\ref{eq:dynamical_system})]. Then, setting all derivatives to zero brings to the following equation that relates $\beta$ with the rest of the parameters:
\begin{equation}
\Delta=\beta^2\pm\sqrt{\mu^2-(1+\alpha_{NL})^2},
\label{eq:Delta}
\end{equation} 
with $\alpha_{NL}=\frac{2\beta\rho}{|d|}\left(1-\frac{\text{ln}(2)}{\beta|d|}\right)$.
From Eq.~(\ref{eq:Delta}), we obtain an expression for $\Delta_{\text{max}}$ imposing $\mu=1+\alpha_{NL}$:

\begin{equation}
\Delta_{\text{max}}=d^2\left(\frac{\mu-1}{2\rho}+\frac{\text{ln}(2)}{d^2}\right)^2.
\label{eq:Delta_max}
\end{equation}

\begin{figure}
    \begin{center}
        \input{figura_4.pgf}
    \end{center}
    \caption{Comparative of $\Delta_{\text{max}}$ calculated using numerical continuation of the stationary states of Eq. (\ref{eq:mean_field}) [solid] and the analytical expression Eq. (\ref{eq:Delta_max}) [dashed]. (a) With a fixed value of $d=6.25$, (b) for constant $\mu=1.1$.}
    \label{fig:comparativa}
\end{figure}
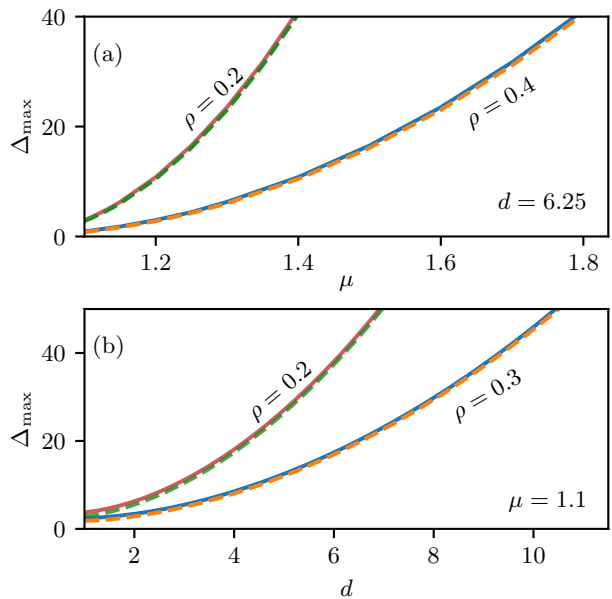

 Figure \ref{fig:comparativa}  shows a comparison between $\Delta_{\text{max}}$ obtained analytically (dashed) and numerically (solid). Note that  $\Delta_{\text{max}}$ increases with $\mu$ and decreases with $\rho$, and for a fixed value of $\mu$, $\Delta_{\text{max}}$ increases with $d$. This result confirms that increasing $d$ extends the interaction region between the soliton and pump; thus, more energy can be transferred to the soliton.

\begin{figure}
    \begin{center}
        \input{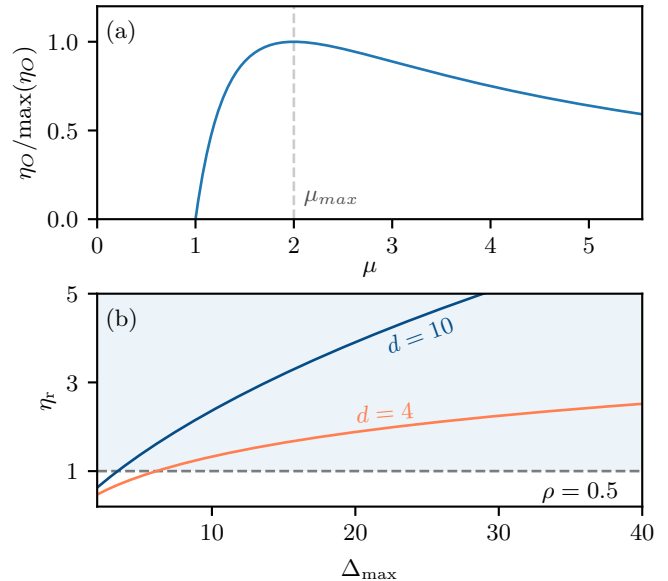}
    \end{center}
    \caption{(a) Normalized conversion efficiency of a PDCS as a function of the pump ($\mu$). (b) Ratio ($\eta_r$) between the maximum efficiencies of a nonlinear coupler ($\eta_O$) and a linear one ($\eta_K$) as a function of $\Delta_{\text{max}}$  as a function of $d$. Dashed line represents $\eta_r=1$, where $\eta_O=\eta_K$.  }
    \label{fig:efficiency}
\end{figure}

The maximum frequency conversion from the pump to the PDCS at fixed $\mu$ is achieved when its peak power and spectral bandwidth are maximum. These conditions are obtained at $\Delta_{\mathrm{max}}$ [given by Eq.~(\ref{eq:Delta_max})]. We define the conversion efficiency as $\eta= \int |E_{\text{sol}}|^2\text{d}\tau/(B_{\text{in}}^2 T_c)$, where $E_{\text{sol}}$ is the dimensional soliton envelope. The maximum conversion efficiency ($\eta$) of the device is: 
\begin{equation}
    \eta_O=\frac{4|\Delta\beta_1|L_2}{\Lambda T_c }\left(\frac{\mu-1}{\mu^2}\right),
    \label{eq:efficiencyO}
\end{equation}
where we have neglected the second term in Eq. (\ref{eq:Delta_max}) since it scales as $d^{-2}$ and we are interested on the large walk-off limit. Eq. (\ref{eq:efficiencyO}) has a maximum at $\mu=2$ [see Fig. \ref{fig:efficiency} (a)]. This maximum corresponds to a pumping intensity of 4 times the intensity at the OPO threshold ($\mu=1$), and it coincides with the maximum conversion of the cw limit~\cite{ebrahimzadeh_optical_chapter}.

It is interesting to compare $\eta_O$ to the maximal conversion obtained with a standard Kerr resonator modeled by the LLE ($\eta_S$).  Considering that both cavities have the same finesse and $T_c$, taking into account that the maximum attainable detuning in the LLE as a function of the normalized pump amplitude ($S$) is $\Delta_{\text{max}}^\text{LLE}=\pi^2 S^2/8$~\cite{Coen_13_universal}, and assuming  that solitons at the same detuning from the LLE  and Eq.~(\ref{eq:mean_field})  are equal, we obtain the following ratio between efficiencies:    

%$S^2=8\gamma L \theta_{in} P_{in}/(\Lambda^3)$

%\begin{equation}
%    \eta_r=\frac{\eta_{OPO}}{\eta_{S}}=\rho\left[\frac{S(\Delta_{\text{max}})}{\mu(\Delta_{\text{max}})}\right]^2.
%\end{equation}

\begin{equation}
    \eta_r=\frac{\eta_{O}}{\eta_{S}}=\rho\left(\frac{\Lambda}{2\theta_{\text{in}}}\right)\left[\frac{S(\Delta_{\text{max}})}{\mu(\Delta_{\text{max}})}\right]^2,
\end{equation}
where $\theta_{\text{in}}$ is the input power coupling of the linear coupler. To calculate $\eta_r$, we consider the minimum $S$ or $\mu$ to achieve a given phase detuning, which corresponds to 
$\Delta_{\text{max}}$.

Figure~\ref{fig:efficiency} (b) shows $\eta_r$ for two different values of $d$  assuming critical coupling ($\theta_{in}=\Lambda/2$).  For high values of $\Delta_{\text{max}}$  (which correspond to high driving powers), a coupler made with a nonlinear crystal is more effective for providing energy. Note that, $\eta_O$ increases with $d$, since PDCS exist for higher values of $\Delta$. This property is also clear from Eq. (\ref{eq:efficiencyO}), where $\eta_O$ is proportional to $|\Delta \beta_1|$. 

In summary, we found analytically the maximum detuning ($\Delta_{\text{max}}$) reached by a PDCS---which corresponds to the shortest PDCS  time duration achievable at fixed driving power---considering pump depletion. Using variational techniques, we  computed the region of the PDCS existence as a function of the normalized walk-off $d$, pump strength $\mu$, and the ratio between second and third-order nonlinearities $\rho$. These results agree with the exact solutions obtained with numerical continuation. Moreover,  using $\Delta_{\text{max}}$ we derived the energy conversion efficiency ($\eta_O$) and its maximum analytically.  We  compared this conversion efficiency to the one of a CS described with the LLE. We showed that driving the cavity with an OPO is more efficient  than with a linear coupler. 
\\
\\
\textbf{Funding} 
European Research Council (grant agreement 757800). Fonds de la Recherche Scientifique-FNRS.
HORIZON EUROPE European Research Council (57800);  H2020 Marie Skłodowska-Curie Actions (101023717).
\\
\\
\textbf{Disclosures} The authors declare no conflict of interest. 
\\
\\
\textbf{Data availability} Data  underlying  the  results  presented  in  this paper are not publicly available at this time but may be obtained from the authors upon reasonable request.

\bibliography{biblioteca2}

\end{document}

%% file: with_crystal.pdf_tex
%% Creator: Inkscape 1.0.2 (1.0.2+r75+1), www.inkscape.org
%% Accompanies image file 'with_crystal.pdf' (pdf, eps, ps)
\begingroup%
  \makeatletter%
  \providecommand\color[2][]{%
    \errmessage{(Inkscape) Color is used for the text in Inkscape, but the package 'color.sty' is not loaded}%
    \renewcommand\color[2][]{}%
  }%
  \providecommand\transparent[1]{%
    \errmessage{(Inkscape) Transparency is used (non-zero) for the text in Inkscape, but the package 'transparent.sty' is not loaded}%
    \renewcommand\transparent[1]{}%
  }%
  \providecommand\rotatebox[2]{#2}%
  \newcommand*\fsize{\dimexpr\f@size pt\relax}%
  \newcommand*\lineheight[1]{\fontsize{\fsize}{#1\fsize}\selectfont}%
  \ifx\svgwidth\undefined%
    \setlength{\unitlength}{246.23823752bp}%
    \ifx\svgscale\undefined%
      \relax%
    \else%
      \setlength{\unitlength}{\unitlength * \real{\svgscale}}%
    \fi%
  \else%
    \setlength{\unitlength}{\svgwidth}%
  \fi%
  \global\let\svgwidth\undefined%
  \global\let\svgscale\undefined%
  \makeatother%
  \begin{picture}(1,0.84825865)%
    \lineheight{1}%
    \setlength\tabcolsep{0pt}%
    \put(0.26554708,0.59852624){\color[rgb]{0,0,0}\makebox(0,0)[lt]{\smash{\begin{tabular}[t]{l}$\color[rgb]{0.8594,0.3984,0.4062}\text{PDCS}$\end{tabular}}}}%
    \put(0.29673062,0.74211676){\makebox(0,0)[lt]{\smash{\begin{tabular}[t]{l}$\color[rgb]{0.3359,0.2666,0.9922}L_2$\end{tabular}}}}%
    \put(0.37970256,0.51842926){\makebox(0,0)[lt]{\smash{\begin{tabular}[t]{l}$\color[rgb]{0.8594,0.3984,0.4062}\omega_0$\end{tabular}}}}%
    \put(0,0){\includegraphics[width=\unitlength,page=1]{with_crystal.pdf}}%
    \put(0.29584123,0.83607534){\color[rgb]{0,0,0}\makebox(0,0)[lt]{\smash{\begin{tabular}[t]{l}$\color[rgb]{0.3359,0.2666,0.9922}\chi^{(2)}$\end{tabular}}}}%
    \put(0.19677001,0.83541412){\color[rgb]{0,0,0}\makebox(0,0)[lt]{\smash{\begin{tabular}[t]{l}$\color[rgb]{0.3984,0.5,0.6133}2\omega_0$\end{tabular}}}}%
    \put(0,0){\includegraphics[width=\unitlength,page=2]{with_crystal.pdf}}%
    \put(0.16765641,0.69332703){\color[rgb]{0,0,0}\makebox(0,0)[lt]{\smash{\begin{tabular}[t]{l}$\chi^{(3)}$\end{tabular}}}}%
    \put(0.16934859,0.61382782){\makebox(0,0)[lt]{\smash{\begin{tabular}[t]{l}$L_3$\end{tabular}}}}%
    \put(0,0){\includegraphics[width=\unitlength,page=3]{with_crystal.pdf}}%
    \put(0.37342311,0.83541412){\color[rgb]{0,0,0}\makebox(0,0)[lt]{\smash{\begin{tabular}[t]{l}$\color[rgb]{0.3984,0.5,0.6133}2\omega_0$\end{tabular}}}}%
    \put(0,0){\includegraphics[width=\unitlength,page=4]{with_crystal.pdf}}%
    \put(0.79370265,0.74978652){\makebox(0,0)[lt]{\smash{\begin{tabular}[t]{l}$\theta$\end{tabular}}}}%
    \put(0,0){\includegraphics[width=\unitlength,page=5]{with_crystal.pdf}}%
    \put(0.66717252,0.66896036){\color[rgb]{0,0,0}\makebox(0,0)[lt]{\smash{\begin{tabular}[t]{l}$\chi^{(3)}$\end{tabular}}}}%
    \put(0.62154469,0.80598081){\color[rgb]{0,0,0}\makebox(0,0)[lt]{\smash{\begin{tabular}[t]{l}$\color[rgb]{0.8594,0.3984,0.4062}\omega_0$\end{tabular}}}}%
    \put(0.85405785,0.57474557){\makebox(0,0)[lt]{\smash{\begin{tabular}[t]{l}$\color[rgb]{0.8594,0.3984,0.4062}\omega_0$\end{tabular}}}}%
    \put(0,0){\includegraphics[width=\unitlength,page=6]{with_crystal.pdf}}%
    \put(0.77935277,0.60507877){\color[rgb]{0,0,0}\makebox(0,0)[lt]{\smash{\begin{tabular}[t]{l}$\color[rgb]{0.8594,0.3984,0.4062}\text{CS}$\end{tabular}}}}%
    \put(0,0){\includegraphics[width=\unitlength,page=7]{with_crystal.pdf}}%
    \put(0.03654925,0.20912756){\rotatebox{90}{\makebox(0,0)[lt]{\smash{\begin{tabular}[t]{l}$|A|_{\text{max}}^2$\end{tabular}}}}}%
    \put(0.77296951,0.00785311){\makebox(0,0)[lt]{\smash{\begin{tabular}[t]{l}$\Delta$\end{tabular}}}}%
    \put(0.2985176,0.00785311){\makebox(0,0)[lt]{\smash{\begin{tabular}[t]{l}$\Delta$\end{tabular}}}}%
    \put(0.05696572,0.79241488){\makebox(0,0)[lt]{\smash{\begin{tabular}[t]{l}(a)\end{tabular}}}}%
    \put(0.54429797,0.79241458){\makebox(0,0)[lt]{\smash{\begin{tabular}[t]{l}(b)\end{tabular}}}}%
    \put(0,0){\includegraphics[width=\unitlength,page=8]{with_crystal.pdf}}%
    \put(0.65233833,0.0414874){\makebox(0,0)[t]{\lineheight{1.25}\smash{\begin{tabular}[t]{c}0\end{tabular}}}}%
    \put(0,0){\includegraphics[width=\unitlength,page=9]{with_crystal.pdf}}%
    \put(0.78217399,0.0414874){\makebox(0,0)[t]{\lineheight{1.25}\smash{\begin{tabular}[t]{c}5\end{tabular}}}}%
    \put(0,0){\includegraphics[width=\unitlength,page=10]{with_crystal.pdf}}%
    \put(0.91200964,0.0414874){\makebox(0,0)[t]{\lineheight{1.25}\smash{\begin{tabular}[t]{c}10\end{tabular}}}}%
    \put(0,0){\includegraphics[width=\unitlength,page=11]{with_crystal.pdf}}%
    \put(0.30937626,0.0414874){\makebox(0,0)[t]{\lineheight{1.25}\smash{\begin{tabular}[t]{c}5\end{tabular}}}}%
    \put(0,0){\includegraphics[width=\unitlength,page=12]{with_crystal.pdf}}%
    \put(0.43921189,0.0414874){\makebox(0,0)[t]{\lineheight{1.25}\smash{\begin{tabular}[t]{c}10\end{tabular}}}}%
    \put(0,0){\includegraphics[width=\unitlength,page=13]{with_crystal.pdf}}%
    \put(0.09917862,0.08380132){\makebox(0,0)[rt]{\lineheight{1.25}\smash{\begin{tabular}[t]{r}0\end{tabular}}}}%
    \put(0,0){\includegraphics[width=\unitlength,page=14]{with_crystal.pdf}}%
    \put(0.09917862,0.23226315){\makebox(0,0)[rt]{\lineheight{1.25}\smash{\begin{tabular}[t]{r}10\end{tabular}}}}%
    \put(0,0){\includegraphics[width=\unitlength,page=15]{with_crystal.pdf}}%
    \put(0.09917862,0.380725){\makebox(0,0)[rt]{\lineheight{1.25}\smash{\begin{tabular}[t]{r}20\end{tabular}}}}%
    \put(0,0){\includegraphics[width=\unitlength,page=16]{with_crystal.pdf}}%
    \put(0.1361574,0.38427317){\makebox(0,0)[lt]{\smash{\begin{tabular}[t]{l}(c)\end{tabular}}}}%
    \put(0.61130613,0.38427813){\makebox(0,0)[lt]{\smash{\begin{tabular}[t]{l}(d)\end{tabular}}}}%
    \put(0,0){\includegraphics[width=\unitlength,page=17]{with_crystal.pdf}}%
    \put(0.19232125,0.0421007){\makebox(0,0)[lt]{\smash{\begin{tabular}[t]{l}$\Delta_0$\end{tabular}}}}%
    \put(0.107038,0.0421007){\makebox(0,0)[lt]{\smash{\begin{tabular}[t]{l}$-\Delta_0$\end{tabular}}}}%
    \put(0,0){\includegraphics[width=\unitlength,page=18]{with_crystal.pdf}}%
    \put(0.41771084,0.33680315){\makebox(0,0)[lt]{\smash{\begin{tabular}[t]{l}$\Delta_{\text{max}}$\end{tabular}}}}%
    \put(0.89725002,0.3368028){\makebox(0,0)[lt]{\smash{\begin{tabular}[t]{l}$\Delta_{\text{max}}$\end{tabular}}}}%
  \end{picture}%
\endgroup%

%% file: figura_3.pgf
%% Creator: Matplotlib, PGF backend
%%
%% To include the figure in your LaTeX document, write
%%   \input{<filename>.pgf}
%%
%% Make sure the required packages are loaded in your preamble
%%   \usepackage{pgf}
%%
%% Figures using additional raster images can only be included by \input if
%% they are in the same directory as the main LaTeX file. For loading figures
%% from other directories you can use the `import` package
%% and then include the figures with
%%   \import{<path to file>}{<filename>.pgf}
%%
%% Matplotlib used the following preamble
%%
\begingroup%
\makeatletter%
\begin{pgfpicture}%
\pgfpathrectangle{\pgfpointorigin}{\pgfqpoint{3.705000in}{3.120000in}}%
\pgfusepath{use as bounding box, clip}%
\begin{pgfscope}%
\pgfsetbuttcap%
\pgfsetmiterjoin%
\definecolor{currentfill}{rgb}{1.000000,1.000000,1.000000}%
\pgfsetfillcolor{currentfill}%
\pgfsetlinewidth{0.000000pt}%
\definecolor{currentstroke}{rgb}{1.000000,1.000000,1.000000}%
\pgfsetstrokecolor{currentstroke}%
\pgfsetdash{}{0pt}%
\pgfpathmoveto{\pgfqpoint{0.000000in}{0.000000in}}%
\pgfpathlineto{\pgfqpoint{3.705000in}{0.000000in}}%
\pgfpathlineto{\pgfqpoint{3.705000in}{3.120000in}}%
\pgfpathlineto{\pgfqpoint{0.000000in}{3.120000in}}%
\pgfpathclose%
\pgfusepath{fill}%
\end{pgfscope}%
\begin{pgfscope}%
\pgfsetbuttcap%
\pgfsetmiterjoin%
\definecolor{currentfill}{rgb}{1.000000,1.000000,1.000000}%
\pgfsetfillcolor{currentfill}%
\pgfsetlinewidth{0.000000pt}%
\definecolor{currentstroke}{rgb}{0.000000,0.000000,0.000000}%
\pgfsetstrokecolor{currentstroke}%
\pgfsetstrokeopacity{0.000000}%
\pgfsetdash{}{0pt}%
\pgfpathmoveto{\pgfqpoint{0.555750in}{1.880758in}}%
\pgfpathlineto{\pgfqpoint{3.297450in}{1.880758in}}%
\pgfpathlineto{\pgfqpoint{3.297450in}{3.057600in}}%
\pgfpathlineto{\pgfqpoint{0.555750in}{3.057600in}}%
\pgfpathclose%
\pgfusepath{fill}%
\end{pgfscope}%
\begin{pgfscope}%
\pgfsetbuttcap%
\pgfsetroundjoin%
\definecolor{currentfill}{rgb}{0.000000,0.000000,0.000000}%
\pgfsetfillcolor{currentfill}%
\pgfsetlinewidth{0.803000pt}%
\definecolor{currentstroke}{rgb}{0.000000,0.000000,0.000000}%
\pgfsetstrokecolor{currentstroke}%
\pgfsetdash{}{0pt}%
\pgfsys@defobject{currentmarker}{\pgfqpoint{0.000000in}{-0.048611in}}{\pgfqpoint{0.000000in}{0.000000in}}{%
\pgfpathmoveto{\pgfqpoint{0.000000in}{0.000000in}}%
\pgfpathlineto{\pgfqpoint{0.000000in}{-0.048611in}}%
\pgfusepath{stroke,fill}%
}%
\begin{pgfscope}%
\pgfsys@transformshift{0.751586in}{1.880758in}%
\pgfsys@useobject{currentmarker}{}%
\end{pgfscope}%
\end{pgfscope}%
\begin{pgfscope}%
\pgftext[x=0.751586in,y=1.783536in,,top]{\rmfamily\fontsize{9.000000}{10.800000}\selectfont \(\displaystyle -6\)}%
\end{pgfscope}%
\begin{pgfscope}%
\pgfsetbuttcap%
\pgfsetroundjoin%
\definecolor{currentfill}{rgb}{0.000000,0.000000,0.000000}%
\pgfsetfillcolor{currentfill}%
\pgfsetlinewidth{0.803000pt}%
\definecolor{currentstroke}{rgb}{0.000000,0.000000,0.000000}%
\pgfsetstrokecolor{currentstroke}%
\pgfsetdash{}{0pt}%
\pgfsys@defobject{currentmarker}{\pgfqpoint{0.000000in}{-0.048611in}}{\pgfqpoint{0.000000in}{0.000000in}}{%
\pgfpathmoveto{\pgfqpoint{0.000000in}{0.000000in}}%
\pgfpathlineto{\pgfqpoint{0.000000in}{-0.048611in}}%
\pgfusepath{stroke,fill}%
}%
\begin{pgfscope}%
\pgfsys@transformshift{1.143257in}{1.880758in}%
\pgfsys@useobject{currentmarker}{}%
\end{pgfscope}%
\end{pgfscope}%
\begin{pgfscope}%
\pgftext[x=1.143257in,y=1.783536in,,top]{\rmfamily\fontsize{9.000000}{10.800000}\selectfont \(\displaystyle -4\)}%
\end{pgfscope}%
\begin{pgfscope}%
\pgfsetbuttcap%
\pgfsetroundjoin%
\definecolor{currentfill}{rgb}{0.000000,0.000000,0.000000}%
\pgfsetfillcolor{currentfill}%
\pgfsetlinewidth{0.803000pt}%
\definecolor{currentstroke}{rgb}{0.000000,0.000000,0.000000}%
\pgfsetstrokecolor{currentstroke}%
\pgfsetdash{}{0pt}%
\pgfsys@defobject{currentmarker}{\pgfqpoint{0.000000in}{-0.048611in}}{\pgfqpoint{0.000000in}{0.000000in}}{%
\pgfpathmoveto{\pgfqpoint{0.000000in}{0.000000in}}%
\pgfpathlineto{\pgfqpoint{0.000000in}{-0.048611in}}%
\pgfusepath{stroke,fill}%
}%
\begin{pgfscope}%
\pgfsys@transformshift{1.534929in}{1.880758in}%
\pgfsys@useobject{currentmarker}{}%
\end{pgfscope}%
\end{pgfscope}%
\begin{pgfscope}%
\pgftext[x=1.534929in,y=1.783536in,,top]{\rmfamily\fontsize{9.000000}{10.800000}\selectfont \(\displaystyle -2\)}%
\end{pgfscope}%
\begin{pgfscope}%
\pgfsetbuttcap%
\pgfsetroundjoin%
\definecolor{currentfill}{rgb}{0.000000,0.000000,0.000000}%
\pgfsetfillcolor{currentfill}%
\pgfsetlinewidth{0.803000pt}%
\definecolor{currentstroke}{rgb}{0.000000,0.000000,0.000000}%
\pgfsetstrokecolor{currentstroke}%
\pgfsetdash{}{0pt}%
\pgfsys@defobject{currentmarker}{\pgfqpoint{0.000000in}{-0.048611in}}{\pgfqpoint{0.000000in}{0.000000in}}{%
\pgfpathmoveto{\pgfqpoint{0.000000in}{0.000000in}}%
\pgfpathlineto{\pgfqpoint{0.000000in}{-0.048611in}}%
\pgfusepath{stroke,fill}%
}%
\begin{pgfscope}%
\pgfsys@transformshift{1.926600in}{1.880758in}%
\pgfsys@useobject{currentmarker}{}%
\end{pgfscope}%
\end{pgfscope}%
\begin{pgfscope}%
\pgftext[x=1.926600in,y=1.783536in,,top]{\rmfamily\fontsize{9.000000}{10.800000}\selectfont \(\displaystyle 0\)}%
\end{pgfscope}%
\begin{pgfscope}%
\pgfsetbuttcap%
\pgfsetroundjoin%
\definecolor{currentfill}{rgb}{0.000000,0.000000,0.000000}%
\pgfsetfillcolor{currentfill}%
\pgfsetlinewidth{0.803000pt}%
\definecolor{currentstroke}{rgb}{0.000000,0.000000,0.000000}%
\pgfsetstrokecolor{currentstroke}%
\pgfsetdash{}{0pt}%
\pgfsys@defobject{currentmarker}{\pgfqpoint{0.000000in}{-0.048611in}}{\pgfqpoint{0.000000in}{0.000000in}}{%
\pgfpathmoveto{\pgfqpoint{0.000000in}{0.000000in}}%
\pgfpathlineto{\pgfqpoint{0.000000in}{-0.048611in}}%
\pgfusepath{stroke,fill}%
}%
\begin{pgfscope}%
\pgfsys@transformshift{2.318271in}{1.880758in}%
\pgfsys@useobject{currentmarker}{}%
\end{pgfscope}%
\end{pgfscope}%
\begin{pgfscope}%
\pgftext[x=2.318271in,y=1.783536in,,top]{\rmfamily\fontsize{9.000000}{10.800000}\selectfont \(\displaystyle 2\)}%
\end{pgfscope}%
\begin{pgfscope}%
\pgfsetbuttcap%
\pgfsetroundjoin%
\definecolor{currentfill}{rgb}{0.000000,0.000000,0.000000}%
\pgfsetfillcolor{currentfill}%
\pgfsetlinewidth{0.803000pt}%
\definecolor{currentstroke}{rgb}{0.000000,0.000000,0.000000}%
\pgfsetstrokecolor{currentstroke}%
\pgfsetdash{}{0pt}%
\pgfsys@defobject{currentmarker}{\pgfqpoint{0.000000in}{-0.048611in}}{\pgfqpoint{0.000000in}{0.000000in}}{%
\pgfpathmoveto{\pgfqpoint{0.000000in}{0.000000in}}%
\pgfpathlineto{\pgfqpoint{0.000000in}{-0.048611in}}%
\pgfusepath{stroke,fill}%
}%
\begin{pgfscope}%
\pgfsys@transformshift{2.709943in}{1.880758in}%
\pgfsys@useobject{currentmarker}{}%
\end{pgfscope}%
\end{pgfscope}%
\begin{pgfscope}%
\pgftext[x=2.709943in,y=1.783536in,,top]{\rmfamily\fontsize{9.000000}{10.800000}\selectfont \(\displaystyle 4\)}%
\end{pgfscope}%
\begin{pgfscope}%
\pgfsetbuttcap%
\pgfsetroundjoin%
\definecolor{currentfill}{rgb}{0.000000,0.000000,0.000000}%
\pgfsetfillcolor{currentfill}%
\pgfsetlinewidth{0.803000pt}%
\definecolor{currentstroke}{rgb}{0.000000,0.000000,0.000000}%
\pgfsetstrokecolor{currentstroke}%
\pgfsetdash{}{0pt}%
\pgfsys@defobject{currentmarker}{\pgfqpoint{0.000000in}{-0.048611in}}{\pgfqpoint{0.000000in}{0.000000in}}{%
\pgfpathmoveto{\pgfqpoint{0.000000in}{0.000000in}}%
\pgfpathlineto{\pgfqpoint{0.000000in}{-0.048611in}}%
\pgfusepath{stroke,fill}%
}%
\begin{pgfscope}%
\pgfsys@transformshift{3.101614in}{1.880758in}%
\pgfsys@useobject{currentmarker}{}%
\end{pgfscope}%
\end{pgfscope}%
\begin{pgfscope}%
\pgftext[x=3.101614in,y=1.783536in,,top]{\rmfamily\fontsize{9.000000}{10.800000}\selectfont \(\displaystyle 6\)}%
\end{pgfscope}%
\begin{pgfscope}%
\pgfsetbuttcap%
\pgfsetroundjoin%
\definecolor{currentfill}{rgb}{0.000000,0.000000,0.000000}%
\pgfsetfillcolor{currentfill}%
\pgfsetlinewidth{0.803000pt}%
\definecolor{currentstroke}{rgb}{0.000000,0.000000,0.000000}%
\pgfsetstrokecolor{currentstroke}%
\pgfsetdash{}{0pt}%
\pgfsys@defobject{currentmarker}{\pgfqpoint{-0.048611in}{0.000000in}}{\pgfqpoint{0.000000in}{0.000000in}}{%
\pgfpathmoveto{\pgfqpoint{0.000000in}{0.000000in}}%
\pgfpathlineto{\pgfqpoint{-0.048611in}{0.000000in}}%
\pgfusepath{stroke,fill}%
}%
\begin{pgfscope}%
\pgfsys@transformshift{0.555750in}{1.983587in}%
\pgfsys@useobject{currentmarker}{}%
\end{pgfscope}%
\end{pgfscope}%
\begin{pgfscope}%
\pgftext[x=0.294370in,y=1.940184in,left,base]{\rmfamily\fontsize{9.000000}{10.800000}\selectfont 0.5}%
\end{pgfscope}%
\begin{pgfscope}%
\pgfsetbuttcap%
\pgfsetroundjoin%
\definecolor{currentfill}{rgb}{0.000000,0.000000,0.000000}%
\pgfsetfillcolor{currentfill}%
\pgfsetlinewidth{0.803000pt}%
\definecolor{currentstroke}{rgb}{0.000000,0.000000,0.000000}%
\pgfsetstrokecolor{currentstroke}%
\pgfsetdash{}{0pt}%
\pgfsys@defobject{currentmarker}{\pgfqpoint{-0.048611in}{0.000000in}}{\pgfqpoint{0.000000in}{0.000000in}}{%
\pgfpathmoveto{\pgfqpoint{0.000000in}{0.000000in}}%
\pgfpathlineto{\pgfqpoint{-0.048611in}{0.000000in}}%
\pgfusepath{stroke,fill}%
}%
\begin{pgfscope}%
\pgfsys@transformshift{0.555750in}{2.663934in}%
\pgfsys@useobject{currentmarker}{}%
\end{pgfscope}%
\end{pgfscope}%
\begin{pgfscope}%
\pgftext[x=0.294370in,y=2.620531in,left,base]{\rmfamily\fontsize{9.000000}{10.800000}\selectfont 1.0}%
\end{pgfscope}%
\begin{pgfscope}%
\pgfsetbuttcap%
\pgfsetroundjoin%
\definecolor{currentfill}{rgb}{0.000000,0.000000,0.000000}%
\pgfsetfillcolor{currentfill}%
\pgfsetlinewidth{0.803000pt}%
\definecolor{currentstroke}{rgb}{0.000000,0.000000,0.000000}%
\pgfsetstrokecolor{currentstroke}%
\pgfsetdash{}{0pt}%
\pgfsys@defobject{currentmarker}{\pgfqpoint{-0.048611in}{0.000000in}}{\pgfqpoint{0.000000in}{0.000000in}}{%
\pgfpathmoveto{\pgfqpoint{0.000000in}{0.000000in}}%
\pgfpathlineto{\pgfqpoint{-0.048611in}{0.000000in}}%
\pgfusepath{stroke,fill}%
}%
\begin{pgfscope}%
\pgfsys@transformshift{0.555750in}{3.004107in}%
\pgfsys@useobject{currentmarker}{}%
\end{pgfscope}%
\end{pgfscope}%
\begin{pgfscope}%
\pgftext[x=0.323257in,y=2.960704in,left,base]{\rmfamily\fontsize{9.000000}{10.800000}\selectfont \(\displaystyle  \mu_0 \)}%
\end{pgfscope}%
\begin{pgfscope}%
\definecolor{textcolor}{rgb}{1.000000,0.000000,0.000000}%
\pgfsetstrokecolor{textcolor}%
\pgfsetfillcolor{textcolor}%
\pgftext[x=0.281580in,y=2.410337in,,bottom,rotate=90.000000]{\color{textcolor}\rmfamily\fontsize{9.000000}{10.800000}\selectfont \(\displaystyle B\)}%
\end{pgfscope}%
\begin{pgfscope}%
\pgfpathrectangle{\pgfqpoint{0.555750in}{1.880758in}}{\pgfqpoint{2.741700in}{1.176842in}} %
\pgfusepath{clip}%
\pgfsetrectcap%
\pgfsetroundjoin%
\pgfsetlinewidth{1.003750pt}%
\definecolor{currentstroke}{rgb}{1.000000,0.000000,0.000000}%
\pgfsetstrokecolor{currentstroke}%
\pgfsetdash{}{0pt}%
\pgfpathmoveto{\pgfqpoint{0.554841in}{3.004107in}}%
\pgfpathlineto{\pgfqpoint{1.617581in}{3.003665in}}%
\pgfpathlineto{\pgfqpoint{1.663007in}{3.002725in}}%
\pgfpathlineto{\pgfqpoint{1.690502in}{3.001354in}}%
\pgfpathlineto{\pgfqpoint{1.710824in}{2.999530in}}%
\pgfpathlineto{\pgfqpoint{1.727561in}{2.997156in}}%
\pgfpathlineto{\pgfqpoint{1.740710in}{2.994463in}}%
\pgfpathlineto{\pgfqpoint{1.752665in}{2.991130in}}%
\pgfpathlineto{\pgfqpoint{1.763423in}{2.987173in}}%
\pgfpathlineto{\pgfqpoint{1.772987in}{2.982675in}}%
\pgfpathlineto{\pgfqpoint{1.781355in}{2.977794in}}%
\pgfpathlineto{\pgfqpoint{1.789723in}{2.971837in}}%
\pgfpathlineto{\pgfqpoint{1.796896in}{2.965711in}}%
\pgfpathlineto{\pgfqpoint{1.804068in}{2.958475in}}%
\pgfpathlineto{\pgfqpoint{1.811241in}{2.949951in}}%
\pgfpathlineto{\pgfqpoint{1.818413in}{2.939940in}}%
\pgfpathlineto{\pgfqpoint{1.825586in}{2.928224in}}%
\pgfpathlineto{\pgfqpoint{1.832759in}{2.914571in}}%
\pgfpathlineto{\pgfqpoint{1.839931in}{2.898738in}}%
\pgfpathlineto{\pgfqpoint{1.847104in}{2.880478in}}%
\pgfpathlineto{\pgfqpoint{1.854276in}{2.859560in}}%
\pgfpathlineto{\pgfqpoint{1.861449in}{2.835772in}}%
\pgfpathlineto{\pgfqpoint{1.868622in}{2.808952in}}%
\pgfpathlineto{\pgfqpoint{1.875794in}{2.779004in}}%
\pgfpathlineto{\pgfqpoint{1.884162in}{2.740108in}}%
\pgfpathlineto{\pgfqpoint{1.892530in}{2.697137in}}%
\pgfpathlineto{\pgfqpoint{1.902094in}{2.643601in}}%
\pgfpathlineto{\pgfqpoint{1.915243in}{2.564231in}}%
\pgfpathlineto{\pgfqpoint{1.947520in}{2.365999in}}%
\pgfpathlineto{\pgfqpoint{1.957084in}{2.313535in}}%
\pgfpathlineto{\pgfqpoint{1.965452in}{2.271929in}}%
\pgfpathlineto{\pgfqpoint{1.973820in}{2.234856in}}%
\pgfpathlineto{\pgfqpoint{1.980992in}{2.206867in}}%
\pgfpathlineto{\pgfqpoint{1.988165in}{2.182388in}}%
\pgfpathlineto{\pgfqpoint{1.995337in}{2.161327in}}%
\pgfpathlineto{\pgfqpoint{2.001315in}{2.146265in}}%
\pgfpathlineto{\pgfqpoint{2.007292in}{2.133332in}}%
\pgfpathlineto{\pgfqpoint{2.013269in}{2.122381in}}%
\pgfpathlineto{\pgfqpoint{2.019246in}{2.113259in}}%
\pgfpathlineto{\pgfqpoint{2.025223in}{2.105806in}}%
\pgfpathlineto{\pgfqpoint{2.031200in}{2.099868in}}%
\pgfpathlineto{\pgfqpoint{2.037177in}{2.095291in}}%
\pgfpathlineto{\pgfqpoint{2.043155in}{2.091932in}}%
\pgfpathlineto{\pgfqpoint{2.049132in}{2.089658in}}%
\pgfpathlineto{\pgfqpoint{2.055109in}{2.088344in}}%
\pgfpathlineto{\pgfqpoint{2.061086in}{2.087878in}}%
\pgfpathlineto{\pgfqpoint{2.068259in}{2.088293in}}%
\pgfpathlineto{\pgfqpoint{2.075431in}{2.089625in}}%
\pgfpathlineto{\pgfqpoint{2.083799in}{2.092157in}}%
\pgfpathlineto{\pgfqpoint{2.093363in}{2.096113in}}%
\pgfpathlineto{\pgfqpoint{2.104122in}{2.101649in}}%
\pgfpathlineto{\pgfqpoint{2.116076in}{2.108844in}}%
\pgfpathlineto{\pgfqpoint{2.130421in}{2.118558in}}%
\pgfpathlineto{\pgfqpoint{2.148353in}{2.131861in}}%
\pgfpathlineto{\pgfqpoint{2.171066in}{2.149899in}}%
\pgfpathlineto{\pgfqpoint{2.202147in}{2.175790in}}%
\pgfpathlineto{\pgfqpoint{2.252355in}{2.218915in}}%
\pgfpathlineto{\pgfqpoint{2.367117in}{2.318983in}}%
\pgfpathlineto{\pgfqpoint{3.001891in}{2.873259in}}%
\pgfpathlineto{\pgfqpoint{3.042536in}{2.907388in}}%
\pgfpathlineto{\pgfqpoint{3.070031in}{2.929341in}}%
\pgfpathlineto{\pgfqpoint{3.090353in}{2.944503in}}%
\pgfpathlineto{\pgfqpoint{3.108285in}{2.956778in}}%
\pgfpathlineto{\pgfqpoint{3.123825in}{2.966339in}}%
\pgfpathlineto{\pgfqpoint{3.138171in}{2.974119in}}%
\pgfpathlineto{\pgfqpoint{3.151320in}{2.980287in}}%
\pgfpathlineto{\pgfqpoint{3.164470in}{2.985511in}}%
\pgfpathlineto{\pgfqpoint{3.177620in}{2.989818in}}%
\pgfpathlineto{\pgfqpoint{3.191965in}{2.993557in}}%
\pgfpathlineto{\pgfqpoint{3.207506in}{2.996629in}}%
\pgfpathlineto{\pgfqpoint{3.224242in}{2.999018in}}%
\pgfpathlineto{\pgfqpoint{3.244564in}{3.000963in}}%
\pgfpathlineto{\pgfqpoint{3.269668in}{3.002398in}}%
\pgfpathlineto{\pgfqpoint{3.298359in}{3.003265in}}%
\pgfpathlineto{\pgfqpoint{3.298359in}{3.003265in}}%
\pgfusepath{stroke}%
\end{pgfscope}%
\begin{pgfscope}%
\pgfpathrectangle{\pgfqpoint{0.555750in}{1.880758in}}{\pgfqpoint{2.741700in}{1.176842in}} %
\pgfusepath{clip}%
\pgfsetbuttcap%
\pgfsetroundjoin%
\pgfsetlinewidth{1.003750pt}%
\definecolor{currentstroke}{rgb}{1.000000,0.000000,0.000000}%
\pgfsetstrokecolor{currentstroke}%
\pgfsetstrokeopacity{0.500000}%
\pgfsetdash{{3.700000pt}{1.600000pt}}{0.000000pt}%
\pgfpathmoveto{\pgfqpoint{0.554841in}{3.004107in}}%
\pgfpathlineto{\pgfqpoint{1.926002in}{3.004107in}}%
\pgfpathlineto{\pgfqpoint{1.927198in}{1.934251in}}%
\pgfpathlineto{\pgfqpoint{3.150125in}{3.003715in}}%
\pgfpathlineto{\pgfqpoint{3.152516in}{3.004107in}}%
\pgfpathlineto{\pgfqpoint{3.298359in}{3.004107in}}%
\pgfpathlineto{\pgfqpoint{3.298359in}{3.004107in}}%
\pgfusepath{stroke}%
\end{pgfscope}%
\begin{pgfscope}%
\pgfsetrectcap%
\pgfsetmiterjoin%
\pgfsetlinewidth{0.803000pt}%
\definecolor{currentstroke}{rgb}{0.000000,0.000000,0.000000}%
\pgfsetstrokecolor{currentstroke}%
\pgfsetdash{}{0pt}%
\pgfpathmoveto{\pgfqpoint{0.555750in}{1.880758in}}%
\pgfpathlineto{\pgfqpoint{0.555750in}{3.057600in}}%
\pgfusepath{stroke}%
\end{pgfscope}%
\begin{pgfscope}%
\pgfsetrectcap%
\pgfsetmiterjoin%
\pgfsetlinewidth{0.803000pt}%
\definecolor{currentstroke}{rgb}{0.000000,0.000000,0.000000}%
\pgfsetstrokecolor{currentstroke}%
\pgfsetdash{}{0pt}%
\pgfpathmoveto{\pgfqpoint{3.297450in}{1.880758in}}%
\pgfpathlineto{\pgfqpoint{3.297450in}{3.057600in}}%
\pgfusepath{stroke}%
\end{pgfscope}%
\begin{pgfscope}%
\pgfsetrectcap%
\pgfsetmiterjoin%
\pgfsetlinewidth{0.803000pt}%
\definecolor{currentstroke}{rgb}{0.000000,0.000000,0.000000}%
\pgfsetstrokecolor{currentstroke}%
\pgfsetdash{}{0pt}%
\pgfpathmoveto{\pgfqpoint{0.555750in}{1.880758in}}%
\pgfpathlineto{\pgfqpoint{3.297450in}{1.880758in}}%
\pgfusepath{stroke}%
\end{pgfscope}%
\begin{pgfscope}%
\pgfsetrectcap%
\pgfsetmiterjoin%
\pgfsetlinewidth{0.803000pt}%
\definecolor{currentstroke}{rgb}{0.000000,0.000000,0.000000}%
\pgfsetstrokecolor{currentstroke}%
\pgfsetdash{}{0pt}%
\pgfpathmoveto{\pgfqpoint{0.555750in}{3.057600in}}%
\pgfpathlineto{\pgfqpoint{3.297450in}{3.057600in}}%
\pgfusepath{stroke}%
\end{pgfscope}%
\begin{pgfscope}%
\pgfsetbuttcap%
\pgfsetroundjoin%
\definecolor{currentfill}{rgb}{0.000000,0.000000,0.000000}%
\pgfsetfillcolor{currentfill}%
\pgfsetlinewidth{0.803000pt}%
\definecolor{currentstroke}{rgb}{0.000000,0.000000,0.000000}%
\pgfsetstrokecolor{currentstroke}%
\pgfsetdash{}{0pt}%
\pgfsys@defobject{currentmarker}{\pgfqpoint{0.000000in}{0.000000in}}{\pgfqpoint{0.048611in}{0.000000in}}{%
\pgfpathmoveto{\pgfqpoint{0.000000in}{0.000000in}}%
\pgfpathlineto{\pgfqpoint{0.048611in}{0.000000in}}%
\pgfusepath{stroke,fill}%
}%
\begin{pgfscope}%
\pgfsys@transformshift{3.297450in}{1.934251in}%
\pgfsys@useobject{currentmarker}{}%
\end{pgfscope}%
\end{pgfscope}%
\begin{pgfscope}%
\pgftext[x=3.394672in,y=1.890848in,left,base]{\rmfamily\fontsize{9.000000}{10.800000}\selectfont \(\displaystyle 0\)}%
\end{pgfscope}%
\begin{pgfscope}%
\pgfsetbuttcap%
\pgfsetroundjoin%
\definecolor{currentfill}{rgb}{0.000000,0.000000,0.000000}%
\pgfsetfillcolor{currentfill}%
\pgfsetlinewidth{0.803000pt}%
\definecolor{currentstroke}{rgb}{0.000000,0.000000,0.000000}%
\pgfsetstrokecolor{currentstroke}%
\pgfsetdash{}{0pt}%
\pgfsys@defobject{currentmarker}{\pgfqpoint{0.000000in}{0.000000in}}{\pgfqpoint{0.048611in}{0.000000in}}{%
\pgfpathmoveto{\pgfqpoint{0.000000in}{0.000000in}}%
\pgfpathlineto{\pgfqpoint{0.048611in}{0.000000in}}%
\pgfusepath{stroke,fill}%
}%
\begin{pgfscope}%
\pgfsys@transformshift{3.297450in}{2.376875in}%
\pgfsys@useobject{currentmarker}{}%
\end{pgfscope}%
\end{pgfscope}%
\begin{pgfscope}%
\pgftext[x=3.394672in,y=2.333472in,left,base]{\rmfamily\fontsize{9.000000}{10.800000}\selectfont \(\displaystyle 5\)}%
\end{pgfscope}%
\begin{pgfscope}%
\pgfsetbuttcap%
\pgfsetroundjoin%
\definecolor{currentfill}{rgb}{0.000000,0.000000,0.000000}%
\pgfsetfillcolor{currentfill}%
\pgfsetlinewidth{0.803000pt}%
\definecolor{currentstroke}{rgb}{0.000000,0.000000,0.000000}%
\pgfsetstrokecolor{currentstroke}%
\pgfsetdash{}{0pt}%
\pgfsys@defobject{currentmarker}{\pgfqpoint{0.000000in}{0.000000in}}{\pgfqpoint{0.048611in}{0.000000in}}{%
\pgfpathmoveto{\pgfqpoint{0.000000in}{0.000000in}}%
\pgfpathlineto{\pgfqpoint{0.048611in}{0.000000in}}%
\pgfusepath{stroke,fill}%
}%
\begin{pgfscope}%
\pgfsys@transformshift{3.297450in}{2.819499in}%
\pgfsys@useobject{currentmarker}{}%
\end{pgfscope}%
\end{pgfscope}%
\begin{pgfscope}%
\pgftext[x=3.394672in,y=2.776096in,left,base]{\rmfamily\fontsize{9.000000}{10.800000}\selectfont \(\displaystyle 10\)}%
\end{pgfscope}%
\begin{pgfscope}%
\definecolor{textcolor}{rgb}{0.023529,0.301961,0.529412}%
\pgfsetstrokecolor{textcolor}%
\pgfsetfillcolor{textcolor}%
\pgftext[x=3.516786in,y=2.410337in,,top,rotate=90.000000]{\color{textcolor}\rmfamily\fontsize{9.000000}{10.800000}\selectfont \(\displaystyle |A|^2\)}%
\end{pgfscope}%
\begin{pgfscope}%
\pgfpathrectangle{\pgfqpoint{0.555750in}{1.880758in}}{\pgfqpoint{2.741700in}{1.176842in}} %
\pgfusepath{clip}%
\pgfsetrectcap%
\pgfsetroundjoin%
\pgfsetlinewidth{1.003750pt}%
\definecolor{currentstroke}{rgb}{0.023529,0.301961,0.529412}%
\pgfsetstrokecolor{currentstroke}%
\pgfsetdash{}{0pt}%
\pgfpathmoveto{\pgfqpoint{0.554841in}{1.934251in}}%
\pgfpathlineto{\pgfqpoint{1.560200in}{1.934684in}}%
\pgfpathlineto{\pgfqpoint{1.605626in}{1.935604in}}%
\pgfpathlineto{\pgfqpoint{1.634317in}{1.937031in}}%
\pgfpathlineto{\pgfqpoint{1.654639in}{1.938877in}}%
\pgfpathlineto{\pgfqpoint{1.671375in}{1.941285in}}%
\pgfpathlineto{\pgfqpoint{1.684525in}{1.944024in}}%
\pgfpathlineto{\pgfqpoint{1.696479in}{1.947424in}}%
\pgfpathlineto{\pgfqpoint{1.707238in}{1.951476in}}%
\pgfpathlineto{\pgfqpoint{1.716802in}{1.956103in}}%
\pgfpathlineto{\pgfqpoint{1.725170in}{1.961147in}}%
\pgfpathlineto{\pgfqpoint{1.733538in}{1.967336in}}%
\pgfpathlineto{\pgfqpoint{1.740710in}{1.973743in}}%
\pgfpathlineto{\pgfqpoint{1.747883in}{1.981361in}}%
\pgfpathlineto{\pgfqpoint{1.755055in}{1.990409in}}%
\pgfpathlineto{\pgfqpoint{1.762228in}{2.001138in}}%
\pgfpathlineto{\pgfqpoint{1.768205in}{2.011569in}}%
\pgfpathlineto{\pgfqpoint{1.774182in}{2.023554in}}%
\pgfpathlineto{\pgfqpoint{1.780160in}{2.037299in}}%
\pgfpathlineto{\pgfqpoint{1.786137in}{2.053032in}}%
\pgfpathlineto{\pgfqpoint{1.792114in}{2.070995in}}%
\pgfpathlineto{\pgfqpoint{1.798091in}{2.091447in}}%
\pgfpathlineto{\pgfqpoint{1.804068in}{2.114660in}}%
\pgfpathlineto{\pgfqpoint{1.811241in}{2.146545in}}%
\pgfpathlineto{\pgfqpoint{1.818413in}{2.183261in}}%
\pgfpathlineto{\pgfqpoint{1.825586in}{2.225225in}}%
\pgfpathlineto{\pgfqpoint{1.832759in}{2.272762in}}%
\pgfpathlineto{\pgfqpoint{1.839931in}{2.326048in}}%
\pgfpathlineto{\pgfqpoint{1.848299in}{2.395400in}}%
\pgfpathlineto{\pgfqpoint{1.857863in}{2.483324in}}%
\pgfpathlineto{\pgfqpoint{1.869817in}{2.602892in}}%
\pgfpathlineto{\pgfqpoint{1.890139in}{2.808002in}}%
\pgfpathlineto{\pgfqpoint{1.897312in}{2.871654in}}%
\pgfpathlineto{\pgfqpoint{1.903289in}{2.917544in}}%
\pgfpathlineto{\pgfqpoint{1.908071in}{2.948312in}}%
\pgfpathlineto{\pgfqpoint{1.912853in}{2.972928in}}%
\pgfpathlineto{\pgfqpoint{1.916439in}{2.986948in}}%
\pgfpathlineto{\pgfqpoint{1.920025in}{2.996913in}}%
\pgfpathlineto{\pgfqpoint{1.922416in}{3.001222in}}%
\pgfpathlineto{\pgfqpoint{1.924807in}{3.003626in}}%
\pgfpathlineto{\pgfqpoint{1.926002in}{3.004107in}}%
\pgfpathlineto{\pgfqpoint{1.927198in}{3.004107in}}%
\pgfpathlineto{\pgfqpoint{1.928393in}{3.003626in}}%
\pgfpathlineto{\pgfqpoint{1.930784in}{3.001222in}}%
\pgfpathlineto{\pgfqpoint{1.933175in}{2.996913in}}%
\pgfpathlineto{\pgfqpoint{1.935566in}{2.990729in}}%
\pgfpathlineto{\pgfqpoint{1.939152in}{2.978040in}}%
\pgfpathlineto{\pgfqpoint{1.942738in}{2.961434in}}%
\pgfpathlineto{\pgfqpoint{1.947520in}{2.933650in}}%
\pgfpathlineto{\pgfqpoint{1.952302in}{2.900098in}}%
\pgfpathlineto{\pgfqpoint{1.958279in}{2.851341in}}%
\pgfpathlineto{\pgfqpoint{1.965452in}{2.785210in}}%
\pgfpathlineto{\pgfqpoint{1.976210in}{2.676936in}}%
\pgfpathlineto{\pgfqpoint{1.995337in}{2.483324in}}%
\pgfpathlineto{\pgfqpoint{2.004901in}{2.395400in}}%
\pgfpathlineto{\pgfqpoint{2.013269in}{2.326048in}}%
\pgfpathlineto{\pgfqpoint{2.021637in}{2.264442in}}%
\pgfpathlineto{\pgfqpoint{2.028809in}{2.217851in}}%
\pgfpathlineto{\pgfqpoint{2.035982in}{2.176787in}}%
\pgfpathlineto{\pgfqpoint{2.043155in}{2.140907in}}%
\pgfpathlineto{\pgfqpoint{2.050327in}{2.109784in}}%
\pgfpathlineto{\pgfqpoint{2.057500in}{2.082950in}}%
\pgfpathlineto{\pgfqpoint{2.064672in}{2.059935in}}%
\pgfpathlineto{\pgfqpoint{2.071845in}{2.040279in}}%
\pgfpathlineto{\pgfqpoint{2.079018in}{2.023554in}}%
\pgfpathlineto{\pgfqpoint{2.086190in}{2.009365in}}%
\pgfpathlineto{\pgfqpoint{2.093363in}{1.997358in}}%
\pgfpathlineto{\pgfqpoint{2.100535in}{1.987220in}}%
\pgfpathlineto{\pgfqpoint{2.107708in}{1.978674in}}%
\pgfpathlineto{\pgfqpoint{2.114881in}{1.971482in}}%
\pgfpathlineto{\pgfqpoint{2.122053in}{1.965437in}}%
\pgfpathlineto{\pgfqpoint{2.130421in}{1.959598in}}%
\pgfpathlineto{\pgfqpoint{2.138789in}{1.954842in}}%
\pgfpathlineto{\pgfqpoint{2.148353in}{1.950480in}}%
\pgfpathlineto{\pgfqpoint{2.159112in}{1.946661in}}%
\pgfpathlineto{\pgfqpoint{2.171066in}{1.943457in}}%
\pgfpathlineto{\pgfqpoint{2.184216in}{1.940877in}}%
\pgfpathlineto{\pgfqpoint{2.199756in}{1.938741in}}%
\pgfpathlineto{\pgfqpoint{2.218883in}{1.937031in}}%
\pgfpathlineto{\pgfqpoint{2.243987in}{1.935732in}}%
\pgfpathlineto{\pgfqpoint{2.279850in}{1.934853in}}%
\pgfpathlineto{\pgfqpoint{2.340817in}{1.934381in}}%
\pgfpathlineto{\pgfqpoint{2.517741in}{1.934252in}}%
\pgfpathlineto{\pgfqpoint{3.298359in}{1.934251in}}%
\pgfpathlineto{\pgfqpoint{3.298359in}{1.934251in}}%
\pgfusepath{stroke}%
\end{pgfscope}%
\begin{pgfscope}%
\pgfsetrectcap%
\pgfsetmiterjoin%
\pgfsetlinewidth{0.803000pt}%
\definecolor{currentstroke}{rgb}{0.000000,0.000000,0.000000}%
\pgfsetstrokecolor{currentstroke}%
\pgfsetdash{}{0pt}%
\pgfpathmoveto{\pgfqpoint{0.555750in}{1.880758in}}%
\pgfpathlineto{\pgfqpoint{0.555750in}{3.057600in}}%
\pgfusepath{stroke}%
\end{pgfscope}%
\begin{pgfscope}%
\pgfsetrectcap%
\pgfsetmiterjoin%
\pgfsetlinewidth{0.803000pt}%
\definecolor{currentstroke}{rgb}{0.000000,0.000000,0.000000}%
\pgfsetstrokecolor{currentstroke}%
\pgfsetdash{}{0pt}%
\pgfpathmoveto{\pgfqpoint{3.297450in}{1.880758in}}%
\pgfpathlineto{\pgfqpoint{3.297450in}{3.057600in}}%
\pgfusepath{stroke}%
\end{pgfscope}%
\begin{pgfscope}%
\pgfsetrectcap%
\pgfsetmiterjoin%
\pgfsetlinewidth{0.803000pt}%
\definecolor{currentstroke}{rgb}{0.000000,0.000000,0.000000}%
\pgfsetstrokecolor{currentstroke}%
\pgfsetdash{}{0pt}%
\pgfpathmoveto{\pgfqpoint{0.555750in}{1.880758in}}%
\pgfpathlineto{\pgfqpoint{3.297450in}{1.880758in}}%
\pgfusepath{stroke}%
\end{pgfscope}%
\begin{pgfscope}%
\pgfsetrectcap%
\pgfsetmiterjoin%
\pgfsetlinewidth{0.803000pt}%
\definecolor{currentstroke}{rgb}{0.000000,0.000000,0.000000}%
\pgfsetstrokecolor{currentstroke}%
\pgfsetdash{}{0pt}%
\pgfpathmoveto{\pgfqpoint{0.555750in}{3.057600in}}%
\pgfpathlineto{\pgfqpoint{3.297450in}{3.057600in}}%
\pgfusepath{stroke}%
\end{pgfscope}%
\begin{pgfscope}%
\pgftext[x=0.596876in,y=2.822232in,left,base]{\rmfamily\fontsize{9.000000}{10.800000}\bfseries\selectfont \(\displaystyle \mathrm{(a)}\)}%
\end{pgfscope}%
\begin{pgfscope}%
\pgfsetbuttcap%
\pgfsetmiterjoin%
\definecolor{currentfill}{rgb}{1.000000,1.000000,1.000000}%
\pgfsetfillcolor{currentfill}%
\pgfsetlinewidth{0.000000pt}%
\definecolor{currentstroke}{rgb}{0.000000,0.000000,0.000000}%
\pgfsetstrokecolor{currentstroke}%
\pgfsetstrokeopacity{0.000000}%
\pgfsetdash{}{0pt}%
\pgfpathmoveto{\pgfqpoint{0.555750in}{0.374400in}}%
\pgfpathlineto{\pgfqpoint{3.297450in}{0.374400in}}%
\pgfpathlineto{\pgfqpoint{3.297450in}{1.551242in}}%
\pgfpathlineto{\pgfqpoint{0.555750in}{1.551242in}}%
\pgfpathclose%
\pgfusepath{fill}%
\end{pgfscope}%
\begin{pgfscope}%
\pgfsys@transformshift{0.556000in}{0.376000in}%
\pgftext[left,bottom]{\pgfimage[interpolate=true,width=2.740000in,height=1.176000in]{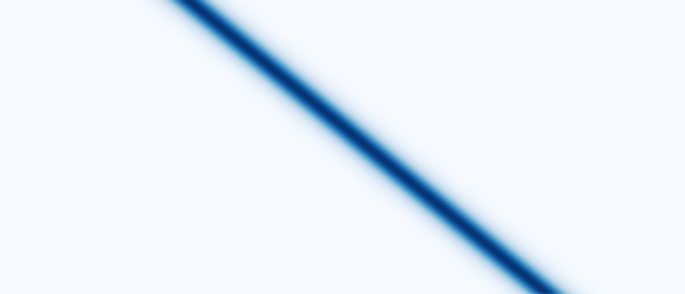}}%
\end{pgfscope}%
\begin{pgfscope}%
\pgfsetbuttcap%
\pgfsetroundjoin%
\definecolor{currentfill}{rgb}{0.000000,0.000000,0.000000}%
\pgfsetfillcolor{currentfill}%
\pgfsetlinewidth{0.803000pt}%
\definecolor{currentstroke}{rgb}{0.000000,0.000000,0.000000}%
\pgfsetstrokecolor{currentstroke}%
\pgfsetdash{}{0pt}%
\pgfsys@defobject{currentmarker}{\pgfqpoint{0.000000in}{-0.048611in}}{\pgfqpoint{0.000000in}{0.000000in}}{%
\pgfpathmoveto{\pgfqpoint{0.000000in}{0.000000in}}%
\pgfpathlineto{\pgfqpoint{0.000000in}{-0.048611in}}%
\pgfusepath{stroke,fill}%
}%
\begin{pgfscope}%
\pgfsys@transformshift{0.555750in}{0.374400in}%
\pgfsys@useobject{currentmarker}{}%
\end{pgfscope}%
\end{pgfscope}%
\begin{pgfscope}%
\pgftext[x=0.555750in,y=0.277178in,,top]{\rmfamily\fontsize{9.000000}{10.800000}\selectfont \(\displaystyle -20\)}%
\end{pgfscope}%
\begin{pgfscope}%
\pgfsetbuttcap%
\pgfsetroundjoin%
\definecolor{currentfill}{rgb}{0.000000,0.000000,0.000000}%
\pgfsetfillcolor{currentfill}%
\pgfsetlinewidth{0.803000pt}%
\definecolor{currentstroke}{rgb}{0.000000,0.000000,0.000000}%
\pgfsetstrokecolor{currentstroke}%
\pgfsetdash{}{0pt}%
\pgfsys@defobject{currentmarker}{\pgfqpoint{0.000000in}{-0.048611in}}{\pgfqpoint{0.000000in}{0.000000in}}{%
\pgfpathmoveto{\pgfqpoint{0.000000in}{0.000000in}}%
\pgfpathlineto{\pgfqpoint{0.000000in}{-0.048611in}}%
\pgfusepath{stroke,fill}%
}%
\begin{pgfscope}%
\pgfsys@transformshift{1.104090in}{0.374400in}%
\pgfsys@useobject{currentmarker}{}%
\end{pgfscope}%
\end{pgfscope}%
\begin{pgfscope}%
\pgftext[x=1.104090in,y=0.277178in,,top]{\rmfamily\fontsize{9.000000}{10.800000}\selectfont \(\displaystyle -15\)}%
\end{pgfscope}%
\begin{pgfscope}%
\pgfsetbuttcap%
\pgfsetroundjoin%
\definecolor{currentfill}{rgb}{0.000000,0.000000,0.000000}%
\pgfsetfillcolor{currentfill}%
\pgfsetlinewidth{0.803000pt}%
\definecolor{currentstroke}{rgb}{0.000000,0.000000,0.000000}%
\pgfsetstrokecolor{currentstroke}%
\pgfsetdash{}{0pt}%
\pgfsys@defobject{currentmarker}{\pgfqpoint{0.000000in}{-0.048611in}}{\pgfqpoint{0.000000in}{0.000000in}}{%
\pgfpathmoveto{\pgfqpoint{0.000000in}{0.000000in}}%
\pgfpathlineto{\pgfqpoint{0.000000in}{-0.048611in}}%
\pgfusepath{stroke,fill}%
}%
\begin{pgfscope}%
\pgfsys@transformshift{1.652430in}{0.374400in}%
\pgfsys@useobject{currentmarker}{}%
\end{pgfscope}%
\end{pgfscope}%
\begin{pgfscope}%
\pgftext[x=1.652430in,y=0.277178in,,top]{\rmfamily\fontsize{9.000000}{10.800000}\selectfont \(\displaystyle -10\)}%
\end{pgfscope}%
\begin{pgfscope}%
\pgfsetbuttcap%
\pgfsetroundjoin%
\definecolor{currentfill}{rgb}{0.000000,0.000000,0.000000}%
\pgfsetfillcolor{currentfill}%
\pgfsetlinewidth{0.803000pt}%
\definecolor{currentstroke}{rgb}{0.000000,0.000000,0.000000}%
\pgfsetstrokecolor{currentstroke}%
\pgfsetdash{}{0pt}%
\pgfsys@defobject{currentmarker}{\pgfqpoint{0.000000in}{-0.048611in}}{\pgfqpoint{0.000000in}{0.000000in}}{%
\pgfpathmoveto{\pgfqpoint{0.000000in}{0.000000in}}%
\pgfpathlineto{\pgfqpoint{0.000000in}{-0.048611in}}%
\pgfusepath{stroke,fill}%
}%
\begin{pgfscope}%
\pgfsys@transformshift{2.200770in}{0.374400in}%
\pgfsys@useobject{currentmarker}{}%
\end{pgfscope}%
\end{pgfscope}%
\begin{pgfscope}%
\pgftext[x=2.200770in,y=0.277178in,,top]{\rmfamily\fontsize{9.000000}{10.800000}\selectfont \(\displaystyle -5\)}%
\end{pgfscope}%
\begin{pgfscope}%
\pgfsetbuttcap%
\pgfsetroundjoin%
\definecolor{currentfill}{rgb}{0.000000,0.000000,0.000000}%
\pgfsetfillcolor{currentfill}%
\pgfsetlinewidth{0.803000pt}%
\definecolor{currentstroke}{rgb}{0.000000,0.000000,0.000000}%
\pgfsetstrokecolor{currentstroke}%
\pgfsetdash{}{0pt}%
\pgfsys@defobject{currentmarker}{\pgfqpoint{0.000000in}{-0.048611in}}{\pgfqpoint{0.000000in}{0.000000in}}{%
\pgfpathmoveto{\pgfqpoint{0.000000in}{0.000000in}}%
\pgfpathlineto{\pgfqpoint{0.000000in}{-0.048611in}}%
\pgfusepath{stroke,fill}%
}%
\begin{pgfscope}%
\pgfsys@transformshift{2.749110in}{0.374400in}%
\pgfsys@useobject{currentmarker}{}%
\end{pgfscope}%
\end{pgfscope}%
\begin{pgfscope}%
\pgftext[x=2.749110in,y=0.277178in,,top]{\rmfamily\fontsize{9.000000}{10.800000}\selectfont \(\displaystyle 0\)}%
\end{pgfscope}%
\begin{pgfscope}%
\pgfsetbuttcap%
\pgfsetroundjoin%
\definecolor{currentfill}{rgb}{0.000000,0.000000,0.000000}%
\pgfsetfillcolor{currentfill}%
\pgfsetlinewidth{0.803000pt}%
\definecolor{currentstroke}{rgb}{0.000000,0.000000,0.000000}%
\pgfsetstrokecolor{currentstroke}%
\pgfsetdash{}{0pt}%
\pgfsys@defobject{currentmarker}{\pgfqpoint{0.000000in}{-0.048611in}}{\pgfqpoint{0.000000in}{0.000000in}}{%
\pgfpathmoveto{\pgfqpoint{0.000000in}{0.000000in}}%
\pgfpathlineto{\pgfqpoint{0.000000in}{-0.048611in}}%
\pgfusepath{stroke,fill}%
}%
\begin{pgfscope}%
\pgfsys@transformshift{3.297450in}{0.374400in}%
\pgfsys@useobject{currentmarker}{}%
\end{pgfscope}%
\end{pgfscope}%
\begin{pgfscope}%
\pgftext[x=3.297450in,y=0.277178in,,top]{\rmfamily\fontsize{9.000000}{10.800000}\selectfont \(\displaystyle 5\)}%
\end{pgfscope}%
\begin{pgfscope}%
\pgftext[x=1.926600in,y=0.110511in,,top]{\rmfamily\fontsize{9.000000}{10.800000}\selectfont \(\displaystyle \tau\)}%
\end{pgfscope}%
\begin{pgfscope}%
\pgfsetbuttcap%
\pgfsetroundjoin%
\definecolor{currentfill}{rgb}{0.000000,0.000000,0.000000}%
\pgfsetfillcolor{currentfill}%
\pgfsetlinewidth{0.803000pt}%
\definecolor{currentstroke}{rgb}{0.000000,0.000000,0.000000}%
\pgfsetstrokecolor{currentstroke}%
\pgfsetdash{}{0pt}%
\pgfsys@defobject{currentmarker}{\pgfqpoint{-0.048611in}{0.000000in}}{\pgfqpoint{0.000000in}{0.000000in}}{%
\pgfpathmoveto{\pgfqpoint{0.000000in}{0.000000in}}%
\pgfpathlineto{\pgfqpoint{-0.048611in}{0.000000in}}%
\pgfusepath{stroke,fill}%
}%
\begin{pgfscope}%
\pgfsys@transformshift{0.555750in}{0.374400in}%
\pgfsys@useobject{currentmarker}{}%
\end{pgfscope}%
\end{pgfscope}%
\begin{pgfscope}%
\pgftext[x=0.394292in,y=0.330997in,left,base]{\rmfamily\fontsize{9.000000}{10.800000}\selectfont \(\displaystyle 0\)}%
\end{pgfscope}%
\begin{pgfscope}%
\pgfsetbuttcap%
\pgfsetroundjoin%
\definecolor{currentfill}{rgb}{0.000000,0.000000,0.000000}%
\pgfsetfillcolor{currentfill}%
\pgfsetlinewidth{0.803000pt}%
\definecolor{currentstroke}{rgb}{0.000000,0.000000,0.000000}%
\pgfsetstrokecolor{currentstroke}%
\pgfsetdash{}{0pt}%
\pgfsys@defobject{currentmarker}{\pgfqpoint{-0.048611in}{0.000000in}}{\pgfqpoint{0.000000in}{0.000000in}}{%
\pgfpathmoveto{\pgfqpoint{0.000000in}{0.000000in}}%
\pgfpathlineto{\pgfqpoint{-0.048611in}{0.000000in}}%
\pgfusepath{stroke,fill}%
}%
\begin{pgfscope}%
\pgfsys@transformshift{0.555750in}{0.847027in}%
\pgfsys@useobject{currentmarker}{}%
\end{pgfscope}%
\end{pgfscope}%
\begin{pgfscope}%
\pgftext[x=0.265821in,y=0.803625in,left,base]{\rmfamily\fontsize{9.000000}{10.800000}\selectfont \(\displaystyle 100\)}%
\end{pgfscope}%
\begin{pgfscope}%
\pgfsetbuttcap%
\pgfsetroundjoin%
\definecolor{currentfill}{rgb}{0.000000,0.000000,0.000000}%
\pgfsetfillcolor{currentfill}%
\pgfsetlinewidth{0.803000pt}%
\definecolor{currentstroke}{rgb}{0.000000,0.000000,0.000000}%
\pgfsetstrokecolor{currentstroke}%
\pgfsetdash{}{0pt}%
\pgfsys@defobject{currentmarker}{\pgfqpoint{-0.048611in}{0.000000in}}{\pgfqpoint{0.000000in}{0.000000in}}{%
\pgfpathmoveto{\pgfqpoint{0.000000in}{0.000000in}}%
\pgfpathlineto{\pgfqpoint{-0.048611in}{0.000000in}}%
\pgfusepath{stroke,fill}%
}%
\begin{pgfscope}%
\pgfsys@transformshift{0.555750in}{1.319655in}%
\pgfsys@useobject{currentmarker}{}%
\end{pgfscope}%
\end{pgfscope}%
\begin{pgfscope}%
\pgftext[x=0.265821in,y=1.276252in,left,base]{\rmfamily\fontsize{9.000000}{10.800000}\selectfont \(\displaystyle 200\)}%
\end{pgfscope}%
\begin{pgfscope}%
\pgftext[x=0.210265in,y=0.962821in,,bottom,rotate=90.000000]{\rmfamily\fontsize{9.000000}{10.800000}\selectfont \(\displaystyle \mathrm{t}\)}%
\end{pgfscope}%
\begin{pgfscope}%
\pgfsetrectcap%
\pgfsetmiterjoin%
\pgfsetlinewidth{0.803000pt}%
\definecolor{currentstroke}{rgb}{0.000000,0.000000,0.000000}%
\pgfsetstrokecolor{currentstroke}%
\pgfsetdash{}{0pt}%
\pgfpathmoveto{\pgfqpoint{0.555750in}{0.374400in}}%
\pgfpathlineto{\pgfqpoint{0.555750in}{1.551242in}}%
\pgfusepath{stroke}%
\end{pgfscope}%
\begin{pgfscope}%
\pgfsetrectcap%
\pgfsetmiterjoin%
\pgfsetlinewidth{0.803000pt}%
\definecolor{currentstroke}{rgb}{0.000000,0.000000,0.000000}%
\pgfsetstrokecolor{currentstroke}%
\pgfsetdash{}{0pt}%
\pgfpathmoveto{\pgfqpoint{3.297450in}{0.374400in}}%
\pgfpathlineto{\pgfqpoint{3.297450in}{1.551242in}}%
\pgfusepath{stroke}%
\end{pgfscope}%
\begin{pgfscope}%
\pgfsetrectcap%
\pgfsetmiterjoin%
\pgfsetlinewidth{0.803000pt}%
\definecolor{currentstroke}{rgb}{0.000000,0.000000,0.000000}%
\pgfsetstrokecolor{currentstroke}%
\pgfsetdash{}{0pt}%
\pgfpathmoveto{\pgfqpoint{0.555750in}{0.374400in}}%
\pgfpathlineto{\pgfqpoint{3.297450in}{0.374400in}}%
\pgfusepath{stroke}%
\end{pgfscope}%
\begin{pgfscope}%
\pgfsetrectcap%
\pgfsetmiterjoin%
\pgfsetlinewidth{0.803000pt}%
\definecolor{currentstroke}{rgb}{0.000000,0.000000,0.000000}%
\pgfsetstrokecolor{currentstroke}%
\pgfsetdash{}{0pt}%
\pgfpathmoveto{\pgfqpoint{0.555750in}{1.551242in}}%
\pgfpathlineto{\pgfqpoint{3.297450in}{1.551242in}}%
\pgfusepath{stroke}%
\end{pgfscope}%
\begin{pgfscope}%
\pgftext[x=0.596876in,y=1.315874in,left,base]{\rmfamily\fontsize{9.000000}{10.800000}\bfseries\selectfont \(\displaystyle \mathrm{(b)}\)}%
\end{pgfscope}%
\end{pgfpicture}%
\makeatother%
\endgroup%

%% file: figura_4.pgf
%% Creator: Matplotlib, PGF backend
%%
%% To include the figure in your LaTeX document, write
%%   \input{<filename>.pgf}
%%
%% Make sure the required packages are loaded in your preamble
%%   \usepackage{pgf}
%%
%% Figures using additional raster images can only be included by \input if
%% they are in the same directory as the main LaTeX file. For loading figures
%% from other directories you can use the `import` package
%% and then include the figures with
%%   \import{<path to file>}{<filename>.pgf}
%%
%% Matplotlib used the following preamble
%%
\begingroup%
\makeatletter%
\begin{pgfpicture}%
\pgfpathrectangle{\pgfpointorigin}{\pgfqpoint{3.705000in}{3.120000in}}%
\pgfusepath{use as bounding box, clip}%
\begin{pgfscope}%
\pgfsetbuttcap%
\pgfsetmiterjoin%
\definecolor{currentfill}{rgb}{1.000000,1.000000,1.000000}%
\pgfsetfillcolor{currentfill}%
\pgfsetlinewidth{0.000000pt}%
\definecolor{currentstroke}{rgb}{1.000000,1.000000,1.000000}%
\pgfsetstrokecolor{currentstroke}%
\pgfsetdash{}{0pt}%
\pgfpathmoveto{\pgfqpoint{0.000000in}{0.000000in}}%
\pgfpathlineto{\pgfqpoint{3.705000in}{0.000000in}}%
\pgfpathlineto{\pgfqpoint{3.705000in}{3.120000in}}%
\pgfpathlineto{\pgfqpoint{0.000000in}{3.120000in}}%
\pgfpathclose%
\pgfusepath{fill}%
\end{pgfscope}%
\begin{pgfscope}%
\pgfsetbuttcap%
\pgfsetmiterjoin%
\definecolor{currentfill}{rgb}{1.000000,1.000000,1.000000}%
\pgfsetfillcolor{currentfill}%
\pgfsetlinewidth{0.000000pt}%
\definecolor{currentstroke}{rgb}{0.000000,0.000000,0.000000}%
\pgfsetstrokecolor{currentstroke}%
\pgfsetstrokeopacity{0.000000}%
\pgfsetdash{}{0pt}%
\pgfpathmoveto{\pgfqpoint{0.555750in}{1.906012in}}%
\pgfpathlineto{\pgfqpoint{3.297450in}{1.906012in}}%
\pgfpathlineto{\pgfqpoint{3.297450in}{3.057600in}}%
\pgfpathlineto{\pgfqpoint{0.555750in}{3.057600in}}%
\pgfpathclose%
\pgfusepath{fill}%
\end{pgfscope}%
\begin{pgfscope}%
\pgfsetbuttcap%
\pgfsetroundjoin%
\definecolor{currentfill}{rgb}{0.000000,0.000000,0.000000}%
\pgfsetfillcolor{currentfill}%
\pgfsetlinewidth{0.803000pt}%
\definecolor{currentstroke}{rgb}{0.000000,0.000000,0.000000}%
\pgfsetstrokecolor{currentstroke}%
\pgfsetdash{}{0pt}%
\pgfsys@defobject{currentmarker}{\pgfqpoint{0.000000in}{-0.048611in}}{\pgfqpoint{0.000000in}{0.000000in}}{%
\pgfpathmoveto{\pgfqpoint{0.000000in}{0.000000in}}%
\pgfpathlineto{\pgfqpoint{0.000000in}{-0.048611in}}%
\pgfusepath{stroke,fill}%
}%
\begin{pgfscope}%
\pgfsys@transformshift{0.928770in}{1.906012in}%
\pgfsys@useobject{currentmarker}{}%
\end{pgfscope}%
\end{pgfscope}%
\begin{pgfscope}%
\definecolor{textcolor}{rgb}{0.000000,0.000000,0.000000}%
\pgfsetstrokecolor{textcolor}%
\pgfsetfillcolor{textcolor}%
\pgftext[x=0.928770in,y=1.808790in,,top]{\color{textcolor}\rmfamily\fontsize{9.000000}{10.800000}\selectfont \(\displaystyle {1.2}\)}%
\end{pgfscope}%
\begin{pgfscope}%
\pgfsetbuttcap%
\pgfsetroundjoin%
\definecolor{currentfill}{rgb}{0.000000,0.000000,0.000000}%
\pgfsetfillcolor{currentfill}%
\pgfsetlinewidth{0.803000pt}%
\definecolor{currentstroke}{rgb}{0.000000,0.000000,0.000000}%
\pgfsetstrokecolor{currentstroke}%
\pgfsetdash{}{0pt}%
\pgfsys@defobject{currentmarker}{\pgfqpoint{0.000000in}{-0.048611in}}{\pgfqpoint{0.000000in}{0.000000in}}{%
\pgfpathmoveto{\pgfqpoint{0.000000in}{0.000000in}}%
\pgfpathlineto{\pgfqpoint{0.000000in}{-0.048611in}}%
\pgfusepath{stroke,fill}%
}%
\begin{pgfscope}%
\pgfsys@transformshift{1.674811in}{1.906012in}%
\pgfsys@useobject{currentmarker}{}%
\end{pgfscope}%
\end{pgfscope}%
\begin{pgfscope}%
\definecolor{textcolor}{rgb}{0.000000,0.000000,0.000000}%
\pgfsetstrokecolor{textcolor}%
\pgfsetfillcolor{textcolor}%
\pgftext[x=1.674811in,y=1.808790in,,top]{\color{textcolor}\rmfamily\fontsize{9.000000}{10.800000}\selectfont \(\displaystyle {1.4}\)}%
\end{pgfscope}%
\begin{pgfscope}%
\pgfsetbuttcap%
\pgfsetroundjoin%
\definecolor{currentfill}{rgb}{0.000000,0.000000,0.000000}%
\pgfsetfillcolor{currentfill}%
\pgfsetlinewidth{0.803000pt}%
\definecolor{currentstroke}{rgb}{0.000000,0.000000,0.000000}%
\pgfsetstrokecolor{currentstroke}%
\pgfsetdash{}{0pt}%
\pgfsys@defobject{currentmarker}{\pgfqpoint{0.000000in}{-0.048611in}}{\pgfqpoint{0.000000in}{0.000000in}}{%
\pgfpathmoveto{\pgfqpoint{0.000000in}{0.000000in}}%
\pgfpathlineto{\pgfqpoint{0.000000in}{-0.048611in}}%
\pgfusepath{stroke,fill}%
}%
\begin{pgfscope}%
\pgfsys@transformshift{2.420852in}{1.906012in}%
\pgfsys@useobject{currentmarker}{}%
\end{pgfscope}%
\end{pgfscope}%
\begin{pgfscope}%
\definecolor{textcolor}{rgb}{0.000000,0.000000,0.000000}%
\pgfsetstrokecolor{textcolor}%
\pgfsetfillcolor{textcolor}%
\pgftext[x=2.420852in,y=1.808790in,,top]{\color{textcolor}\rmfamily\fontsize{9.000000}{10.800000}\selectfont \(\displaystyle {1.6}\)}%
\end{pgfscope}%
\begin{pgfscope}%
\pgfsetbuttcap%
\pgfsetroundjoin%
\definecolor{currentfill}{rgb}{0.000000,0.000000,0.000000}%
\pgfsetfillcolor{currentfill}%
\pgfsetlinewidth{0.803000pt}%
\definecolor{currentstroke}{rgb}{0.000000,0.000000,0.000000}%
\pgfsetstrokecolor{currentstroke}%
\pgfsetdash{}{0pt}%
\pgfsys@defobject{currentmarker}{\pgfqpoint{0.000000in}{-0.048611in}}{\pgfqpoint{0.000000in}{0.000000in}}{%
\pgfpathmoveto{\pgfqpoint{0.000000in}{0.000000in}}%
\pgfpathlineto{\pgfqpoint{0.000000in}{-0.048611in}}%
\pgfusepath{stroke,fill}%
}%
\begin{pgfscope}%
\pgfsys@transformshift{3.166893in}{1.906012in}%
\pgfsys@useobject{currentmarker}{}%
\end{pgfscope}%
\end{pgfscope}%
\begin{pgfscope}%
\definecolor{textcolor}{rgb}{0.000000,0.000000,0.000000}%
\pgfsetstrokecolor{textcolor}%
\pgfsetfillcolor{textcolor}%
\pgftext[x=3.166893in,y=1.808790in,,top]{\color{textcolor}\rmfamily\fontsize{9.000000}{10.800000}\selectfont \(\displaystyle {1.8}\)}%
\end{pgfscope}%
\begin{pgfscope}%
\definecolor{textcolor}{rgb}{0.000000,0.000000,0.000000}%
\pgfsetstrokecolor{textcolor}%
\pgfsetfillcolor{textcolor}%
\pgftext[x=1.926600in,y=1.698726in,,top]{\color{textcolor}\rmfamily\fontsize{9.000000}{10.800000}\selectfont \(\displaystyle \mu\)}%
\end{pgfscope}%
\begin{pgfscope}%
\pgfsetbuttcap%
\pgfsetroundjoin%
\definecolor{currentfill}{rgb}{0.000000,0.000000,0.000000}%
\pgfsetfillcolor{currentfill}%
\pgfsetlinewidth{0.803000pt}%
\definecolor{currentstroke}{rgb}{0.000000,0.000000,0.000000}%
\pgfsetstrokecolor{currentstroke}%
\pgfsetdash{}{0pt}%
\pgfsys@defobject{currentmarker}{\pgfqpoint{-0.048611in}{0.000000in}}{\pgfqpoint{-0.000000in}{0.000000in}}{%
\pgfpathmoveto{\pgfqpoint{-0.000000in}{0.000000in}}%
\pgfpathlineto{\pgfqpoint{-0.048611in}{0.000000in}}%
\pgfusepath{stroke,fill}%
}%
\begin{pgfscope}%
\pgfsys@transformshift{0.555750in}{1.906012in}%
\pgfsys@useobject{currentmarker}{}%
\end{pgfscope}%
\end{pgfscope}%
\begin{pgfscope}%
\definecolor{textcolor}{rgb}{0.000000,0.000000,0.000000}%
\pgfsetstrokecolor{textcolor}%
\pgfsetfillcolor{textcolor}%
\pgftext[x=0.394292in, y=1.862609in, left, base]{\color{textcolor}\rmfamily\fontsize{9.000000}{10.800000}\selectfont \(\displaystyle {0}\)}%
\end{pgfscope}%
\begin{pgfscope}%
\pgfsetbuttcap%
\pgfsetroundjoin%
\definecolor{currentfill}{rgb}{0.000000,0.000000,0.000000}%
\pgfsetfillcolor{currentfill}%
\pgfsetlinewidth{0.803000pt}%
\definecolor{currentstroke}{rgb}{0.000000,0.000000,0.000000}%
\pgfsetstrokecolor{currentstroke}%
\pgfsetdash{}{0pt}%
\pgfsys@defobject{currentmarker}{\pgfqpoint{-0.048611in}{0.000000in}}{\pgfqpoint{-0.000000in}{0.000000in}}{%
\pgfpathmoveto{\pgfqpoint{-0.000000in}{0.000000in}}%
\pgfpathlineto{\pgfqpoint{-0.048611in}{0.000000in}}%
\pgfusepath{stroke,fill}%
}%
\begin{pgfscope}%
\pgfsys@transformshift{0.555750in}{2.481806in}%
\pgfsys@useobject{currentmarker}{}%
\end{pgfscope}%
\end{pgfscope}%
\begin{pgfscope}%
\definecolor{textcolor}{rgb}{0.000000,0.000000,0.000000}%
\pgfsetstrokecolor{textcolor}%
\pgfsetfillcolor{textcolor}%
\pgftext[x=0.330056in, y=2.438403in, left, base]{\color{textcolor}\rmfamily\fontsize{9.000000}{10.800000}\selectfont \(\displaystyle {20}\)}%
\end{pgfscope}%
\begin{pgfscope}%
\pgfsetbuttcap%
\pgfsetroundjoin%
\definecolor{currentfill}{rgb}{0.000000,0.000000,0.000000}%
\pgfsetfillcolor{currentfill}%
\pgfsetlinewidth{0.803000pt}%
\definecolor{currentstroke}{rgb}{0.000000,0.000000,0.000000}%
\pgfsetstrokecolor{currentstroke}%
\pgfsetdash{}{0pt}%
\pgfsys@defobject{currentmarker}{\pgfqpoint{-0.048611in}{0.000000in}}{\pgfqpoint{-0.000000in}{0.000000in}}{%
\pgfpathmoveto{\pgfqpoint{-0.000000in}{0.000000in}}%
\pgfpathlineto{\pgfqpoint{-0.048611in}{0.000000in}}%
\pgfusepath{stroke,fill}%
}%
\begin{pgfscope}%
\pgfsys@transformshift{0.555750in}{3.057600in}%
\pgfsys@useobject{currentmarker}{}%
\end{pgfscope}%
\end{pgfscope}%
\begin{pgfscope}%
\definecolor{textcolor}{rgb}{0.000000,0.000000,0.000000}%
\pgfsetstrokecolor{textcolor}%
\pgfsetfillcolor{textcolor}%
\pgftext[x=0.330056in, y=3.014197in, left, base]{\color{textcolor}\rmfamily\fontsize{9.000000}{10.800000}\selectfont \(\displaystyle {40}\)}%
\end{pgfscope}%
\begin{pgfscope}%
\definecolor{textcolor}{rgb}{0.000000,0.000000,0.000000}%
\pgfsetstrokecolor{textcolor}%
\pgfsetfillcolor{textcolor}%
\pgftext[x=0.274501in,y=2.481806in,,bottom,rotate=90.000000]{\color{textcolor}\rmfamily\fontsize{9.000000}{10.800000}\selectfont \(\displaystyle \Delta_{\mathrm{max}}\)}%
\end{pgfscope}%
\begin{pgfscope}%
\pgfpathrectangle{\pgfqpoint{0.555750in}{1.906012in}}{\pgfqpoint{2.741700in}{1.151588in}}%
\pgfusepath{clip}%
\pgfsetrectcap%
\pgfsetroundjoin%
\pgfsetlinewidth{1.505625pt}%
\definecolor{currentstroke}{rgb}{0.803922,0.360784,0.360784}%
\pgfsetstrokecolor{currentstroke}%
\pgfsetdash{}{0pt}%
\pgfpathmoveto{\pgfqpoint{0.555750in}{1.991665in}}%
\pgfpathlineto{\pgfqpoint{0.593052in}{2.007841in}}%
\pgfpathlineto{\pgfqpoint{0.649005in}{2.034770in}}%
\pgfpathlineto{\pgfqpoint{0.667656in}{2.044342in}}%
\pgfpathlineto{\pgfqpoint{0.704958in}{2.064844in}}%
\pgfpathlineto{\pgfqpoint{0.742260in}{2.086659in}}%
\pgfpathlineto{\pgfqpoint{0.928770in}{2.216790in}}%
\pgfpathlineto{\pgfqpoint{1.115281in}{2.382070in}}%
\pgfpathlineto{\pgfqpoint{1.301791in}{2.582642in}}%
\pgfpathlineto{\pgfqpoint{1.488301in}{2.818419in}}%
\pgfpathlineto{\pgfqpoint{1.661936in}{3.071489in}}%
\pgfusepath{stroke}%
\end{pgfscope}%
\begin{pgfscope}%
\pgfpathrectangle{\pgfqpoint{0.555750in}{1.906012in}}{\pgfqpoint{2.741700in}{1.151588in}}%
\pgfusepath{clip}%
\pgfsetbuttcap%
\pgfsetroundjoin%
\pgfsetlinewidth{1.505625pt}%
\definecolor{currentstroke}{rgb}{0.133333,0.545098,0.133333}%
\pgfsetstrokecolor{currentstroke}%
\pgfsetdash{{5.550000pt}{2.400000pt}}{0.000000pt}%
\pgfpathmoveto{\pgfqpoint{0.555750in}{1.986631in}}%
\pgfpathlineto{\pgfqpoint{0.593052in}{2.002389in}}%
\pgfpathlineto{\pgfqpoint{0.649005in}{2.028662in}}%
\pgfpathlineto{\pgfqpoint{0.667656in}{2.038123in}}%
\pgfpathlineto{\pgfqpoint{0.704958in}{2.058098in}}%
\pgfpathlineto{\pgfqpoint{0.742260in}{2.079479in}}%
\pgfpathlineto{\pgfqpoint{0.928770in}{2.207471in}}%
\pgfpathlineto{\pgfqpoint{1.115281in}{2.370606in}}%
\pgfpathlineto{\pgfqpoint{1.301791in}{2.568886in}}%
\pgfpathlineto{\pgfqpoint{1.488301in}{2.802308in}}%
\pgfpathlineto{\pgfqpoint{1.674811in}{3.070875in}}%
\pgfusepath{stroke}%
\end{pgfscope}%
\begin{pgfscope}%
\pgfpathrectangle{\pgfqpoint{0.555750in}{1.906012in}}{\pgfqpoint{2.741700in}{1.151588in}}%
\pgfusepath{clip}%
\pgfsetrectcap%
\pgfsetroundjoin%
\pgfsetlinewidth{1.505625pt}%
\definecolor{currentstroke}{rgb}{0.121569,0.466667,0.705882}%
\pgfsetstrokecolor{currentstroke}%
\pgfsetdash{}{0pt}%
\pgfpathmoveto{\pgfqpoint{0.555750in}{1.932323in}}%
\pgfpathlineto{\pgfqpoint{0.742260in}{1.957743in}}%
\pgfpathlineto{\pgfqpoint{0.928770in}{1.991947in}}%
\pgfpathlineto{\pgfqpoint{1.115281in}{2.033322in}}%
\pgfpathlineto{\pgfqpoint{1.301791in}{2.086907in}}%
\pgfpathlineto{\pgfqpoint{1.674811in}{2.217019in}}%
\pgfpathlineto{\pgfqpoint{2.047832in}{2.382350in}}%
\pgfpathlineto{\pgfqpoint{2.420852in}{2.582811in}}%
\pgfpathlineto{\pgfqpoint{2.793872in}{2.818605in}}%
\pgfpathlineto{\pgfqpoint{3.140761in}{3.071489in}}%
\pgfusepath{stroke}%
\end{pgfscope}%
\begin{pgfscope}%
\pgfpathrectangle{\pgfqpoint{0.555750in}{1.906012in}}{\pgfqpoint{2.741700in}{1.151588in}}%
\pgfusepath{clip}%
\pgfsetbuttcap%
\pgfsetroundjoin%
\pgfsetlinewidth{1.505625pt}%
\definecolor{currentstroke}{rgb}{1.000000,0.498039,0.054902}%
\pgfsetstrokecolor{currentstroke}%
\pgfsetdash{{5.550000pt}{2.400000pt}}{0.000000pt}%
\pgfpathmoveto{\pgfqpoint{0.555750in}{1.928927in}}%
\pgfpathlineto{\pgfqpoint{0.742260in}{1.953386in}}%
\pgfpathlineto{\pgfqpoint{0.928770in}{1.986631in}}%
\pgfpathlineto{\pgfqpoint{1.115281in}{2.028662in}}%
\pgfpathlineto{\pgfqpoint{1.301791in}{2.079479in}}%
\pgfpathlineto{\pgfqpoint{1.674811in}{2.207471in}}%
\pgfpathlineto{\pgfqpoint{2.047832in}{2.370606in}}%
\pgfpathlineto{\pgfqpoint{2.420852in}{2.568886in}}%
\pgfpathlineto{\pgfqpoint{2.793872in}{2.802308in}}%
\pgfpathlineto{\pgfqpoint{3.166893in}{3.070875in}}%
\pgfusepath{stroke}%
\end{pgfscope}%
\begin{pgfscope}%
\pgfsetrectcap%
\pgfsetmiterjoin%
\pgfsetlinewidth{0.803000pt}%
\definecolor{currentstroke}{rgb}{0.000000,0.000000,0.000000}%
\pgfsetstrokecolor{currentstroke}%
\pgfsetdash{}{0pt}%
\pgfpathmoveto{\pgfqpoint{0.555750in}{1.906012in}}%
\pgfpathlineto{\pgfqpoint{0.555750in}{3.057600in}}%
\pgfusepath{stroke}%
\end{pgfscope}%
\begin{pgfscope}%
\pgfsetrectcap%
\pgfsetmiterjoin%
\pgfsetlinewidth{0.803000pt}%
\definecolor{currentstroke}{rgb}{0.000000,0.000000,0.000000}%
\pgfsetstrokecolor{currentstroke}%
\pgfsetdash{}{0pt}%
\pgfpathmoveto{\pgfqpoint{3.297450in}{1.906012in}}%
\pgfpathlineto{\pgfqpoint{3.297450in}{3.057600in}}%
\pgfusepath{stroke}%
\end{pgfscope}%
\begin{pgfscope}%
\pgfsetrectcap%
\pgfsetmiterjoin%
\pgfsetlinewidth{0.803000pt}%
\definecolor{currentstroke}{rgb}{0.000000,0.000000,0.000000}%
\pgfsetstrokecolor{currentstroke}%
\pgfsetdash{}{0pt}%
\pgfpathmoveto{\pgfqpoint{0.555750in}{1.906012in}}%
\pgfpathlineto{\pgfqpoint{3.297450in}{1.906012in}}%
\pgfusepath{stroke}%
\end{pgfscope}%
\begin{pgfscope}%
\pgfsetrectcap%
\pgfsetmiterjoin%
\pgfsetlinewidth{0.803000pt}%
\definecolor{currentstroke}{rgb}{0.000000,0.000000,0.000000}%
\pgfsetstrokecolor{currentstroke}%
\pgfsetdash{}{0pt}%
\pgfpathmoveto{\pgfqpoint{0.555750in}{3.057600in}}%
\pgfpathlineto{\pgfqpoint{3.297450in}{3.057600in}}%
\pgfusepath{stroke}%
\end{pgfscope}%
\begin{pgfscope}%
\definecolor{textcolor}{rgb}{0.000000,0.000000,0.000000}%
\pgfsetstrokecolor{textcolor}%
\pgfsetfillcolor{textcolor}%
\pgftext[x=1.098761in, y=2.475563in, left, base,rotate=42.000000]{\color{textcolor}\rmfamily\fontsize{9.000000}{10.800000}\selectfont \(\displaystyle \rho=0.2\)}%
\end{pgfscope}%
\begin{pgfscope}%
\definecolor{textcolor}{rgb}{0.000000,0.000000,0.000000}%
\pgfsetstrokecolor{textcolor}%
\pgfsetfillcolor{textcolor}%
\pgftext[x=2.576161in, y=2.478550in, left, base,rotate=30.000000]{\color{textcolor}\rmfamily\fontsize{9.000000}{10.800000}\selectfont \(\displaystyle \rho=0.4\)}%
\end{pgfscope}%
\begin{pgfscope}%
\definecolor{textcolor}{rgb}{0.000000,0.000000,0.000000}%
\pgfsetstrokecolor{textcolor}%
\pgfsetfillcolor{textcolor}%
\pgftext[x=0.596875in,y=2.827282in,left,base]{\color{textcolor}\rmfamily\fontsize{9.000000}{10.800000}\bfseries\selectfont \(\displaystyle \mathrm{(a)}\)}%
\end{pgfscope}%
\begin{pgfscope}%
\definecolor{textcolor}{rgb}{0.000000,0.000000,0.000000}%
\pgfsetstrokecolor{textcolor}%
\pgfsetfillcolor{textcolor}%
\pgftext[x=2.719268in,y=2.049961in,left,base]{\color{textcolor}\rmfamily\fontsize{9.000000}{10.800000}\selectfont \(\displaystyle d=6.25\)}%
\end{pgfscope}%
\begin{pgfscope}%
\pgfsetbuttcap%
\pgfsetmiterjoin%
\definecolor{currentfill}{rgb}{1.000000,1.000000,1.000000}%
\pgfsetfillcolor{currentfill}%
\pgfsetlinewidth{0.000000pt}%
\definecolor{currentstroke}{rgb}{0.000000,0.000000,0.000000}%
\pgfsetstrokecolor{currentstroke}%
\pgfsetstrokeopacity{0.000000}%
\pgfsetdash{}{0pt}%
\pgfpathmoveto{\pgfqpoint{0.555750in}{0.374400in}}%
\pgfpathlineto{\pgfqpoint{3.297450in}{0.374400in}}%
\pgfpathlineto{\pgfqpoint{3.297450in}{1.525988in}}%
\pgfpathlineto{\pgfqpoint{0.555750in}{1.525988in}}%
\pgfpathclose%
\pgfusepath{fill}%
\end{pgfscope}%
\begin{pgfscope}%
\pgfsetbuttcap%
\pgfsetroundjoin%
\definecolor{currentfill}{rgb}{0.000000,0.000000,0.000000}%
\pgfsetfillcolor{currentfill}%
\pgfsetlinewidth{0.803000pt}%
\definecolor{currentstroke}{rgb}{0.000000,0.000000,0.000000}%
\pgfsetstrokecolor{currentstroke}%
\pgfsetdash{}{0pt}%
\pgfsys@defobject{currentmarker}{\pgfqpoint{0.000000in}{-0.048611in}}{\pgfqpoint{0.000000in}{0.000000in}}{%
\pgfpathmoveto{\pgfqpoint{0.000000in}{0.000000in}}%
\pgfpathlineto{\pgfqpoint{0.000000in}{-0.048611in}}%
\pgfusepath{stroke,fill}%
}%
\begin{pgfscope}%
\pgfsys@transformshift{0.816864in}{0.374400in}%
\pgfsys@useobject{currentmarker}{}%
\end{pgfscope}%
\end{pgfscope}%
\begin{pgfscope}%
\definecolor{textcolor}{rgb}{0.000000,0.000000,0.000000}%
\pgfsetstrokecolor{textcolor}%
\pgfsetfillcolor{textcolor}%
\pgftext[x=0.816864in,y=0.277178in,,top]{\color{textcolor}\rmfamily\fontsize{9.000000}{10.800000}\selectfont \(\displaystyle {2}\)}%
\end{pgfscope}%
\begin{pgfscope}%
\pgfsetbuttcap%
\pgfsetroundjoin%
\definecolor{currentfill}{rgb}{0.000000,0.000000,0.000000}%
\pgfsetfillcolor{currentfill}%
\pgfsetlinewidth{0.803000pt}%
\definecolor{currentstroke}{rgb}{0.000000,0.000000,0.000000}%
\pgfsetstrokecolor{currentstroke}%
\pgfsetdash{}{0pt}%
\pgfsys@defobject{currentmarker}{\pgfqpoint{0.000000in}{-0.048611in}}{\pgfqpoint{0.000000in}{0.000000in}}{%
\pgfpathmoveto{\pgfqpoint{0.000000in}{0.000000in}}%
\pgfpathlineto{\pgfqpoint{0.000000in}{-0.048611in}}%
\pgfusepath{stroke,fill}%
}%
\begin{pgfscope}%
\pgfsys@transformshift{1.339093in}{0.374400in}%
\pgfsys@useobject{currentmarker}{}%
\end{pgfscope}%
\end{pgfscope}%
\begin{pgfscope}%
\definecolor{textcolor}{rgb}{0.000000,0.000000,0.000000}%
\pgfsetstrokecolor{textcolor}%
\pgfsetfillcolor{textcolor}%
\pgftext[x=1.339093in,y=0.277178in,,top]{\color{textcolor}\rmfamily\fontsize{9.000000}{10.800000}\selectfont \(\displaystyle {4}\)}%
\end{pgfscope}%
\begin{pgfscope}%
\pgfsetbuttcap%
\pgfsetroundjoin%
\definecolor{currentfill}{rgb}{0.000000,0.000000,0.000000}%
\pgfsetfillcolor{currentfill}%
\pgfsetlinewidth{0.803000pt}%
\definecolor{currentstroke}{rgb}{0.000000,0.000000,0.000000}%
\pgfsetstrokecolor{currentstroke}%
\pgfsetdash{}{0pt}%
\pgfsys@defobject{currentmarker}{\pgfqpoint{0.000000in}{-0.048611in}}{\pgfqpoint{0.000000in}{0.000000in}}{%
\pgfpathmoveto{\pgfqpoint{0.000000in}{0.000000in}}%
\pgfpathlineto{\pgfqpoint{0.000000in}{-0.048611in}}%
\pgfusepath{stroke,fill}%
}%
\begin{pgfscope}%
\pgfsys@transformshift{1.861321in}{0.374400in}%
\pgfsys@useobject{currentmarker}{}%
\end{pgfscope}%
\end{pgfscope}%
\begin{pgfscope}%
\definecolor{textcolor}{rgb}{0.000000,0.000000,0.000000}%
\pgfsetstrokecolor{textcolor}%
\pgfsetfillcolor{textcolor}%
\pgftext[x=1.861321in,y=0.277178in,,top]{\color{textcolor}\rmfamily\fontsize{9.000000}{10.800000}\selectfont \(\displaystyle {6}\)}%
\end{pgfscope}%
\begin{pgfscope}%
\pgfsetbuttcap%
\pgfsetroundjoin%
\definecolor{currentfill}{rgb}{0.000000,0.000000,0.000000}%
\pgfsetfillcolor{currentfill}%
\pgfsetlinewidth{0.803000pt}%
\definecolor{currentstroke}{rgb}{0.000000,0.000000,0.000000}%
\pgfsetstrokecolor{currentstroke}%
\pgfsetdash{}{0pt}%
\pgfsys@defobject{currentmarker}{\pgfqpoint{0.000000in}{-0.048611in}}{\pgfqpoint{0.000000in}{0.000000in}}{%
\pgfpathmoveto{\pgfqpoint{0.000000in}{0.000000in}}%
\pgfpathlineto{\pgfqpoint{0.000000in}{-0.048611in}}%
\pgfusepath{stroke,fill}%
}%
\begin{pgfscope}%
\pgfsys@transformshift{2.383550in}{0.374400in}%
\pgfsys@useobject{currentmarker}{}%
\end{pgfscope}%
\end{pgfscope}%
\begin{pgfscope}%
\definecolor{textcolor}{rgb}{0.000000,0.000000,0.000000}%
\pgfsetstrokecolor{textcolor}%
\pgfsetfillcolor{textcolor}%
\pgftext[x=2.383550in,y=0.277178in,,top]{\color{textcolor}\rmfamily\fontsize{9.000000}{10.800000}\selectfont \(\displaystyle {8}\)}%
\end{pgfscope}%
\begin{pgfscope}%
\pgfsetbuttcap%
\pgfsetroundjoin%
\definecolor{currentfill}{rgb}{0.000000,0.000000,0.000000}%
\pgfsetfillcolor{currentfill}%
\pgfsetlinewidth{0.803000pt}%
\definecolor{currentstroke}{rgb}{0.000000,0.000000,0.000000}%
\pgfsetstrokecolor{currentstroke}%
\pgfsetdash{}{0pt}%
\pgfsys@defobject{currentmarker}{\pgfqpoint{0.000000in}{-0.048611in}}{\pgfqpoint{0.000000in}{0.000000in}}{%
\pgfpathmoveto{\pgfqpoint{0.000000in}{0.000000in}}%
\pgfpathlineto{\pgfqpoint{0.000000in}{-0.048611in}}%
\pgfusepath{stroke,fill}%
}%
\begin{pgfscope}%
\pgfsys@transformshift{2.905779in}{0.374400in}%
\pgfsys@useobject{currentmarker}{}%
\end{pgfscope}%
\end{pgfscope}%
\begin{pgfscope}%
\definecolor{textcolor}{rgb}{0.000000,0.000000,0.000000}%
\pgfsetstrokecolor{textcolor}%
\pgfsetfillcolor{textcolor}%
\pgftext[x=2.905779in,y=0.277178in,,top]{\color{textcolor}\rmfamily\fontsize{9.000000}{10.800000}\selectfont \(\displaystyle {10}\)}%
\end{pgfscope}%
\begin{pgfscope}%
\definecolor{textcolor}{rgb}{0.000000,0.000000,0.000000}%
\pgfsetstrokecolor{textcolor}%
\pgfsetfillcolor{textcolor}%
\pgftext[x=1.926600in,y=0.109535in,,top]{\color{textcolor}\rmfamily\fontsize{9.000000}{10.800000}\selectfont \(\displaystyle d\)}%
\end{pgfscope}%
\begin{pgfscope}%
\pgfsetbuttcap%
\pgfsetroundjoin%
\definecolor{currentfill}{rgb}{0.000000,0.000000,0.000000}%
\pgfsetfillcolor{currentfill}%
\pgfsetlinewidth{0.803000pt}%
\definecolor{currentstroke}{rgb}{0.000000,0.000000,0.000000}%
\pgfsetstrokecolor{currentstroke}%
\pgfsetdash{}{0pt}%
\pgfsys@defobject{currentmarker}{\pgfqpoint{-0.048611in}{0.000000in}}{\pgfqpoint{-0.000000in}{0.000000in}}{%
\pgfpathmoveto{\pgfqpoint{-0.000000in}{0.000000in}}%
\pgfpathlineto{\pgfqpoint{-0.048611in}{0.000000in}}%
\pgfusepath{stroke,fill}%
}%
\begin{pgfscope}%
\pgfsys@transformshift{0.555750in}{0.374400in}%
\pgfsys@useobject{currentmarker}{}%
\end{pgfscope}%
\end{pgfscope}%
\begin{pgfscope}%
\definecolor{textcolor}{rgb}{0.000000,0.000000,0.000000}%
\pgfsetstrokecolor{textcolor}%
\pgfsetfillcolor{textcolor}%
\pgftext[x=0.394292in, y=0.330997in, left, base]{\color{textcolor}\rmfamily\fontsize{9.000000}{10.800000}\selectfont \(\displaystyle {0}\)}%
\end{pgfscope}%
\begin{pgfscope}%
\pgfsetbuttcap%
\pgfsetroundjoin%
\definecolor{currentfill}{rgb}{0.000000,0.000000,0.000000}%
\pgfsetfillcolor{currentfill}%
\pgfsetlinewidth{0.803000pt}%
\definecolor{currentstroke}{rgb}{0.000000,0.000000,0.000000}%
\pgfsetstrokecolor{currentstroke}%
\pgfsetdash{}{0pt}%
\pgfsys@defobject{currentmarker}{\pgfqpoint{-0.048611in}{0.000000in}}{\pgfqpoint{-0.000000in}{0.000000in}}{%
\pgfpathmoveto{\pgfqpoint{-0.000000in}{0.000000in}}%
\pgfpathlineto{\pgfqpoint{-0.048611in}{0.000000in}}%
\pgfusepath{stroke,fill}%
}%
\begin{pgfscope}%
\pgfsys@transformshift{0.555750in}{0.835035in}%
\pgfsys@useobject{currentmarker}{}%
\end{pgfscope}%
\end{pgfscope}%
\begin{pgfscope}%
\definecolor{textcolor}{rgb}{0.000000,0.000000,0.000000}%
\pgfsetstrokecolor{textcolor}%
\pgfsetfillcolor{textcolor}%
\pgftext[x=0.330056in, y=0.791632in, left, base]{\color{textcolor}\rmfamily\fontsize{9.000000}{10.800000}\selectfont \(\displaystyle {20}\)}%
\end{pgfscope}%
\begin{pgfscope}%
\pgfsetbuttcap%
\pgfsetroundjoin%
\definecolor{currentfill}{rgb}{0.000000,0.000000,0.000000}%
\pgfsetfillcolor{currentfill}%
\pgfsetlinewidth{0.803000pt}%
\definecolor{currentstroke}{rgb}{0.000000,0.000000,0.000000}%
\pgfsetstrokecolor{currentstroke}%
\pgfsetdash{}{0pt}%
\pgfsys@defobject{currentmarker}{\pgfqpoint{-0.048611in}{0.000000in}}{\pgfqpoint{-0.000000in}{0.000000in}}{%
\pgfpathmoveto{\pgfqpoint{-0.000000in}{0.000000in}}%
\pgfpathlineto{\pgfqpoint{-0.048611in}{0.000000in}}%
\pgfusepath{stroke,fill}%
}%
\begin{pgfscope}%
\pgfsys@transformshift{0.555750in}{1.295670in}%
\pgfsys@useobject{currentmarker}{}%
\end{pgfscope}%
\end{pgfscope}%
\begin{pgfscope}%
\definecolor{textcolor}{rgb}{0.000000,0.000000,0.000000}%
\pgfsetstrokecolor{textcolor}%
\pgfsetfillcolor{textcolor}%
\pgftext[x=0.330056in, y=1.252268in, left, base]{\color{textcolor}\rmfamily\fontsize{9.000000}{10.800000}\selectfont \(\displaystyle {40}\)}%
\end{pgfscope}%
\begin{pgfscope}%
\definecolor{textcolor}{rgb}{0.000000,0.000000,0.000000}%
\pgfsetstrokecolor{textcolor}%
\pgfsetfillcolor{textcolor}%
\pgftext[x=0.274501in,y=0.950194in,,bottom,rotate=90.000000]{\color{textcolor}\rmfamily\fontsize{9.000000}{10.800000}\selectfont \(\displaystyle \Delta_{\mathrm{max}}\)}%
\end{pgfscope}%
\begin{pgfscope}%
\pgfpathrectangle{\pgfqpoint{0.555750in}{0.374400in}}{\pgfqpoint{2.741700in}{1.151588in}}%
\pgfusepath{clip}%
\pgfsetrectcap%
\pgfsetroundjoin%
\pgfsetlinewidth{1.505625pt}%
\definecolor{currentstroke}{rgb}{0.121569,0.466667,0.705882}%
\pgfsetstrokecolor{currentstroke}%
\pgfsetdash{}{0pt}%
\pgfpathmoveto{\pgfqpoint{0.555750in}{0.429822in}}%
\pgfpathlineto{\pgfqpoint{0.644618in}{0.434731in}}%
\pgfpathlineto{\pgfqpoint{0.736099in}{0.442935in}}%
\pgfpathlineto{\pgfqpoint{0.827581in}{0.454246in}}%
\pgfpathlineto{\pgfqpoint{0.921676in}{0.469028in}}%
\pgfpathlineto{\pgfqpoint{1.015771in}{0.486907in}}%
\pgfpathlineto{\pgfqpoint{1.112480in}{0.508408in}}%
\pgfpathlineto{\pgfqpoint{1.211803in}{0.533688in}}%
\pgfpathlineto{\pgfqpoint{1.311126in}{0.562120in}}%
\pgfpathlineto{\pgfqpoint{1.413062in}{0.594489in}}%
\pgfpathlineto{\pgfqpoint{1.514999in}{0.630041in}}%
\pgfpathlineto{\pgfqpoint{1.619549in}{0.669784in}}%
\pgfpathlineto{\pgfqpoint{1.724099in}{0.712835in}}%
\pgfpathlineto{\pgfqpoint{1.828649in}{0.759205in}}%
\pgfpathlineto{\pgfqpoint{1.933200in}{0.808908in}}%
\pgfpathlineto{\pgfqpoint{2.040364in}{0.863302in}}%
\pgfpathlineto{\pgfqpoint{2.147528in}{0.921160in}}%
\pgfpathlineto{\pgfqpoint{2.254692in}{0.982471in}}%
\pgfpathlineto{\pgfqpoint{2.364470in}{1.048876in}}%
\pgfpathlineto{\pgfqpoint{2.474247in}{1.118953in}}%
\pgfpathlineto{\pgfqpoint{2.584025in}{1.192745in}}%
\pgfpathlineto{\pgfqpoint{2.693803in}{1.270306in}}%
\pgfpathlineto{\pgfqpoint{2.800967in}{1.349771in}}%
\pgfpathlineto{\pgfqpoint{2.908131in}{1.433124in}}%
\pgfpathlineto{\pgfqpoint{3.012681in}{1.518366in}}%
\pgfpathlineto{\pgfqpoint{3.038314in}{1.539877in}}%
\pgfpathlineto{\pgfqpoint{3.038314in}{1.539877in}}%
\pgfusepath{stroke}%
\end{pgfscope}%
\begin{pgfscope}%
\pgfpathrectangle{\pgfqpoint{0.555750in}{0.374400in}}{\pgfqpoint{2.741700in}{1.151588in}}%
\pgfusepath{clip}%
\pgfsetbuttcap%
\pgfsetroundjoin%
\pgfsetlinewidth{1.505625pt}%
\definecolor{currentstroke}{rgb}{1.000000,0.498039,0.054902}%
\pgfsetstrokecolor{currentstroke}%
\pgfsetdash{{5.550000pt}{2.400000pt}}{0.000000pt}%
\pgfpathmoveto{\pgfqpoint{0.555750in}{0.416988in}}%
\pgfpathlineto{\pgfqpoint{0.610639in}{0.418233in}}%
\pgfpathlineto{\pgfqpoint{0.678597in}{0.422937in}}%
\pgfpathlineto{\pgfqpoint{0.759623in}{0.431637in}}%
\pgfpathlineto{\pgfqpoint{0.851104in}{0.444613in}}%
\pgfpathlineto{\pgfqpoint{0.947813in}{0.461508in}}%
\pgfpathlineto{\pgfqpoint{1.047136in}{0.482034in}}%
\pgfpathlineto{\pgfqpoint{1.149073in}{0.506328in}}%
\pgfpathlineto{\pgfqpoint{1.251009in}{0.533832in}}%
\pgfpathlineto{\pgfqpoint{1.352946in}{0.564514in}}%
\pgfpathlineto{\pgfqpoint{1.457496in}{0.599264in}}%
\pgfpathlineto{\pgfqpoint{1.562046in}{0.637323in}}%
\pgfpathlineto{\pgfqpoint{1.666597in}{0.678683in}}%
\pgfpathlineto{\pgfqpoint{1.771147in}{0.723340in}}%
\pgfpathlineto{\pgfqpoint{1.878311in}{0.772531in}}%
\pgfpathlineto{\pgfqpoint{1.985475in}{0.825178in}}%
\pgfpathlineto{\pgfqpoint{2.092639in}{0.881280in}}%
\pgfpathlineto{\pgfqpoint{2.199803in}{0.940835in}}%
\pgfpathlineto{\pgfqpoint{2.309581in}{1.005422in}}%
\pgfpathlineto{\pgfqpoint{2.419358in}{1.073632in}}%
\pgfpathlineto{\pgfqpoint{2.529136in}{1.145462in}}%
\pgfpathlineto{\pgfqpoint{2.641528in}{1.222754in}}%
\pgfpathlineto{\pgfqpoint{2.753919in}{1.303841in}}%
\pgfpathlineto{\pgfqpoint{2.866311in}{1.388723in}}%
\pgfpathlineto{\pgfqpoint{2.981316in}{1.479506in}}%
\pgfpathlineto{\pgfqpoint{3.055142in}{1.539877in}}%
\pgfpathlineto{\pgfqpoint{3.055142in}{1.539877in}}%
\pgfusepath{stroke}%
\end{pgfscope}%
\begin{pgfscope}%
\pgfpathrectangle{\pgfqpoint{0.555750in}{0.374400in}}{\pgfqpoint{2.741700in}{1.151588in}}%
\pgfusepath{clip}%
\pgfsetrectcap%
\pgfsetroundjoin%
\pgfsetlinewidth{1.505625pt}%
\definecolor{currentstroke}{rgb}{0.803922,0.360784,0.360784}%
\pgfsetstrokecolor{currentstroke}%
\pgfsetdash{}{0pt}%
\pgfpathmoveto{\pgfqpoint{0.555750in}{0.460404in}}%
\pgfpathlineto{\pgfqpoint{0.616128in}{0.468885in}}%
\pgfpathlineto{\pgfqpoint{0.676506in}{0.480449in}}%
\pgfpathlineto{\pgfqpoint{0.738713in}{0.495515in}}%
\pgfpathlineto{\pgfqpoint{0.802750in}{0.514281in}}%
\pgfpathlineto{\pgfqpoint{0.866787in}{0.536270in}}%
\pgfpathlineto{\pgfqpoint{0.932654in}{0.562157in}}%
\pgfpathlineto{\pgfqpoint{1.000350in}{0.592125in}}%
\pgfpathlineto{\pgfqpoint{1.069876in}{0.626343in}}%
\pgfpathlineto{\pgfqpoint{1.141231in}{0.664985in}}%
\pgfpathlineto{\pgfqpoint{1.212587in}{0.707156in}}%
\pgfpathlineto{\pgfqpoint{1.285772in}{0.754052in}}%
\pgfpathlineto{\pgfqpoint{1.358957in}{0.804615in}}%
\pgfpathlineto{\pgfqpoint{1.433972in}{0.860239in}}%
\pgfpathlineto{\pgfqpoint{1.508987in}{0.919714in}}%
\pgfpathlineto{\pgfqpoint{1.585831in}{0.984646in}}%
\pgfpathlineto{\pgfqpoint{1.662676in}{1.053644in}}%
\pgfpathlineto{\pgfqpoint{1.741350in}{1.128513in}}%
\pgfpathlineto{\pgfqpoint{1.820024in}{1.207677in}}%
\pgfpathlineto{\pgfqpoint{1.900528in}{1.293152in}}%
\pgfpathlineto{\pgfqpoint{1.979202in}{1.381218in}}%
\pgfpathlineto{\pgfqpoint{2.056046in}{1.471862in}}%
\pgfpathlineto{\pgfqpoint{2.111179in}{1.539877in}}%
\pgfpathlineto{\pgfqpoint{2.111179in}{1.539877in}}%
\pgfusepath{stroke}%
\end{pgfscope}%
\begin{pgfscope}%
\pgfpathrectangle{\pgfqpoint{0.555750in}{0.374400in}}{\pgfqpoint{2.741700in}{1.151588in}}%
\pgfusepath{clip}%
\pgfsetbuttcap%
\pgfsetroundjoin%
\pgfsetlinewidth{1.505625pt}%
\definecolor{currentstroke}{rgb}{0.133333,0.545098,0.133333}%
\pgfsetstrokecolor{currentstroke}%
\pgfsetstrokeopacity{0.800000}%
\pgfsetdash{{5.550000pt}{2.400000pt}}{0.000000pt}%
\pgfpathmoveto{\pgfqpoint{0.555750in}{0.440426in}}%
\pgfpathlineto{\pgfqpoint{0.601491in}{0.446149in}}%
\pgfpathlineto{\pgfqpoint{0.654550in}{0.455912in}}%
\pgfpathlineto{\pgfqpoint{0.713098in}{0.469791in}}%
\pgfpathlineto{\pgfqpoint{0.777135in}{0.488212in}}%
\pgfpathlineto{\pgfqpoint{0.843002in}{0.510418in}}%
\pgfpathlineto{\pgfqpoint{0.910698in}{0.536525in}}%
\pgfpathlineto{\pgfqpoint{0.980224in}{0.566713in}}%
\pgfpathlineto{\pgfqpoint{1.049750in}{0.600267in}}%
\pgfpathlineto{\pgfqpoint{1.121106in}{0.638172in}}%
\pgfpathlineto{\pgfqpoint{1.192461in}{0.679566in}}%
\pgfpathlineto{\pgfqpoint{1.265646in}{0.725632in}}%
\pgfpathlineto{\pgfqpoint{1.338831in}{0.775344in}}%
\pgfpathlineto{\pgfqpoint{1.413846in}{0.830076in}}%
\pgfpathlineto{\pgfqpoint{1.488861in}{0.888626in}}%
\pgfpathlineto{\pgfqpoint{1.565706in}{0.952560in}}%
\pgfpathlineto{\pgfqpoint{1.642550in}{1.020493in}}%
\pgfpathlineto{\pgfqpoint{1.721224in}{1.094184in}}%
\pgfpathlineto{\pgfqpoint{1.801728in}{1.173926in}}%
\pgfpathlineto{\pgfqpoint{1.882231in}{1.258052in}}%
\pgfpathlineto{\pgfqpoint{1.964565in}{1.348623in}}%
\pgfpathlineto{\pgfqpoint{2.046898in}{1.443778in}}%
\pgfpathlineto{\pgfqpoint{2.126292in}{1.539877in}}%
\pgfpathlineto{\pgfqpoint{2.126292in}{1.539877in}}%
\pgfusepath{stroke}%
\end{pgfscope}%
\begin{pgfscope}%
\pgfsetrectcap%
\pgfsetmiterjoin%
\pgfsetlinewidth{0.803000pt}%
\definecolor{currentstroke}{rgb}{0.000000,0.000000,0.000000}%
\pgfsetstrokecolor{currentstroke}%
\pgfsetdash{}{0pt}%
\pgfpathmoveto{\pgfqpoint{0.555750in}{0.374400in}}%
\pgfpathlineto{\pgfqpoint{0.555750in}{1.525988in}}%
\pgfusepath{stroke}%
\end{pgfscope}%
\begin{pgfscope}%
\pgfsetrectcap%
\pgfsetmiterjoin%
\pgfsetlinewidth{0.803000pt}%
\definecolor{currentstroke}{rgb}{0.000000,0.000000,0.000000}%
\pgfsetstrokecolor{currentstroke}%
\pgfsetdash{}{0pt}%
\pgfpathmoveto{\pgfqpoint{3.297450in}{0.374400in}}%
\pgfpathlineto{\pgfqpoint{3.297450in}{1.525988in}}%
\pgfusepath{stroke}%
\end{pgfscope}%
\begin{pgfscope}%
\pgfsetrectcap%
\pgfsetmiterjoin%
\pgfsetlinewidth{0.803000pt}%
\definecolor{currentstroke}{rgb}{0.000000,0.000000,0.000000}%
\pgfsetstrokecolor{currentstroke}%
\pgfsetdash{}{0pt}%
\pgfpathmoveto{\pgfqpoint{0.555750in}{0.374400in}}%
\pgfpathlineto{\pgfqpoint{3.297450in}{0.374400in}}%
\pgfusepath{stroke}%
\end{pgfscope}%
\begin{pgfscope}%
\pgfsetrectcap%
\pgfsetmiterjoin%
\pgfsetlinewidth{0.803000pt}%
\definecolor{currentstroke}{rgb}{0.000000,0.000000,0.000000}%
\pgfsetstrokecolor{currentstroke}%
\pgfsetdash{}{0pt}%
\pgfpathmoveto{\pgfqpoint{0.555750in}{1.525988in}}%
\pgfpathlineto{\pgfqpoint{3.297450in}{1.525988in}}%
\pgfusepath{stroke}%
\end{pgfscope}%
\begin{pgfscope}%
\definecolor{textcolor}{rgb}{0.000000,0.000000,0.000000}%
\pgfsetstrokecolor{textcolor}%
\pgfsetfillcolor{textcolor}%
\pgftext[x=2.505287in, y=0.946938in, left, base,rotate=30.000000]{\color{textcolor}\rmfamily\fontsize{9.000000}{10.800000}\selectfont \(\displaystyle \rho=0.3\)}%
\end{pgfscope}%
\begin{pgfscope}%
\definecolor{textcolor}{rgb}{0.000000,0.000000,0.000000}%
\pgfsetstrokecolor{textcolor}%
\pgfsetfillcolor{textcolor}%
\pgftext[x=1.447113in, y=0.944508in, left, base,rotate=40.000000]{\color{textcolor}\rmfamily\fontsize{9.000000}{10.800000}\selectfont \(\displaystyle \rho=0.2\)}%
\end{pgfscope}%
\begin{pgfscope}%
\definecolor{textcolor}{rgb}{0.000000,0.000000,0.000000}%
\pgfsetstrokecolor{textcolor}%
\pgfsetfillcolor{textcolor}%
\pgftext[x=0.596875in,y=1.295670in,left,base]{\color{textcolor}\rmfamily\fontsize{9.000000}{10.800000}\bfseries\selectfont \(\displaystyle \mathrm{(b)}\)}%
\end{pgfscope}%
\begin{pgfscope}%
\definecolor{textcolor}{rgb}{0.000000,0.000000,0.000000}%
\pgfsetstrokecolor{textcolor}%
\pgfsetfillcolor{textcolor}%
\pgftext[x=2.775221in,y=0.489559in,left,base]{\color{textcolor}\rmfamily\fontsize{9.000000}{10.800000}\selectfont \(\displaystyle \mu=1.1\)}%
\end{pgfscope}%
\end{pgfpicture}%
\makeatother%
\endgroup%